\documentclass{article}

\usepackage{neurips_2025}

\usepackage[utf8]{inputenc} 
\usepackage[T1]{fontenc}    
\usepackage[colorlinks=true,
            linkcolor=darkblue,
            citecolor=darkblue,
            urlcolor=darkblue]{hyperref}
\usepackage{xcolor}  
\definecolor{darkblue}{rgb}{0.0, 0.0, 0.5}
\usepackage{url}            
\usepackage{booktabs}       
\usepackage{amsfonts}       
\usepackage{nicefrac}       
\usepackage{microtype}      
\usepackage{xcolor}         

\usepackage{etoolbox}

\newcommand{\method}{\textsc{MR-FlowDPO}\xspace}
\newcommand{\methSmall}{\textsc{MR-FlowDPO-400M}\xspace}
\newcommand{\methMedium}{\textsc{MR-FlowDPO-1B}\xspace}
\newcommand{\paretomethod}{MRSD\xspace}
\newcommand{\pmrtiny}[1]{\fontsize{5}{6}\selectfont$\pm$#1}

\usepackage{microtype}
\usepackage{graphicx}
\usepackage{subfigure}
\usepackage{booktabs} 

\usepackage{amsmath}
\usepackage{amssymb}
\usepackage{mathtools}
\usepackage{amsthm}
\usepackage{xspace}

\usepackage{algorithm}

\usepackage{algpseudocode}

\usepackage[capitalize,noabbrev]{cleveref}

\usepackage{titlesec}

\title{\method: Multi-Reward Direct Preference Optimization for Flow-Matching Text-to-Music Generation}

  \author{\thanks{Work was done as part of Alon's internship at FAIR.}~Alon Ziv$^{1,2}$, Sanyuan Chen$^{1}$, Andros Tjandra$^{1}$, Yossi Adi$^{1,2}$, Wei-Ning Hsu$^{1}$, Bowen Shi$^{1}$\\\\
$^{1}$FAIR Team, Meta MSL\\
$^{2}$The Hebrew University of Jerusalem\\
\texttt{alonzi@cs.huji.ac.il}
}

\begin{document}

\maketitle

\begin{abstract}
A key challenge in music generation models is their lack of direct alignment with human preferences, as music evaluation is inherently subjective and varies widely across individuals. We introduce \method, a novel approach that enhances flow-matching-based music generation models - a major class of modern music generative models, using Direct Preference Optimization (DPO) with multiple musical rewards. The rewards are crafted to assess music quality across three key dimensions: text alignment, audio production quality, and semantic consistency, utilizing scalable off-the-shelf models for each reward prediction. We employ these rewards in two ways: (i) By constructing preference data for DPO and (ii) by integrating the rewards into text prompting. To address the ambiguity in musicality evaluation, we propose a novel scoring mechanism leveraging semantic self-supervised representations, which significantly improves the rhythmic stability of generated music. We conduct an extensive evaluation using a variety of music-specific objective metrics as well as a human study. Results show that \method significantly enhances overall music generation quality and is consistently preferred over highly competitive baselines in terms of audio quality, text alignment, and musicality. Our code is publicly available.~\footnote{\url{https://github.com/lonzi/mrflow_dpo/}} Samples are provided in our demo page.~\footnote{\url{https://lonzi.github.io/mr_flowdpo_demopage/}}
\end{abstract}

\section{Introduction}
\label{sec:introduction}

\begin{figure*}[ht!]
    \centering
    \includegraphics[width=\textwidth]{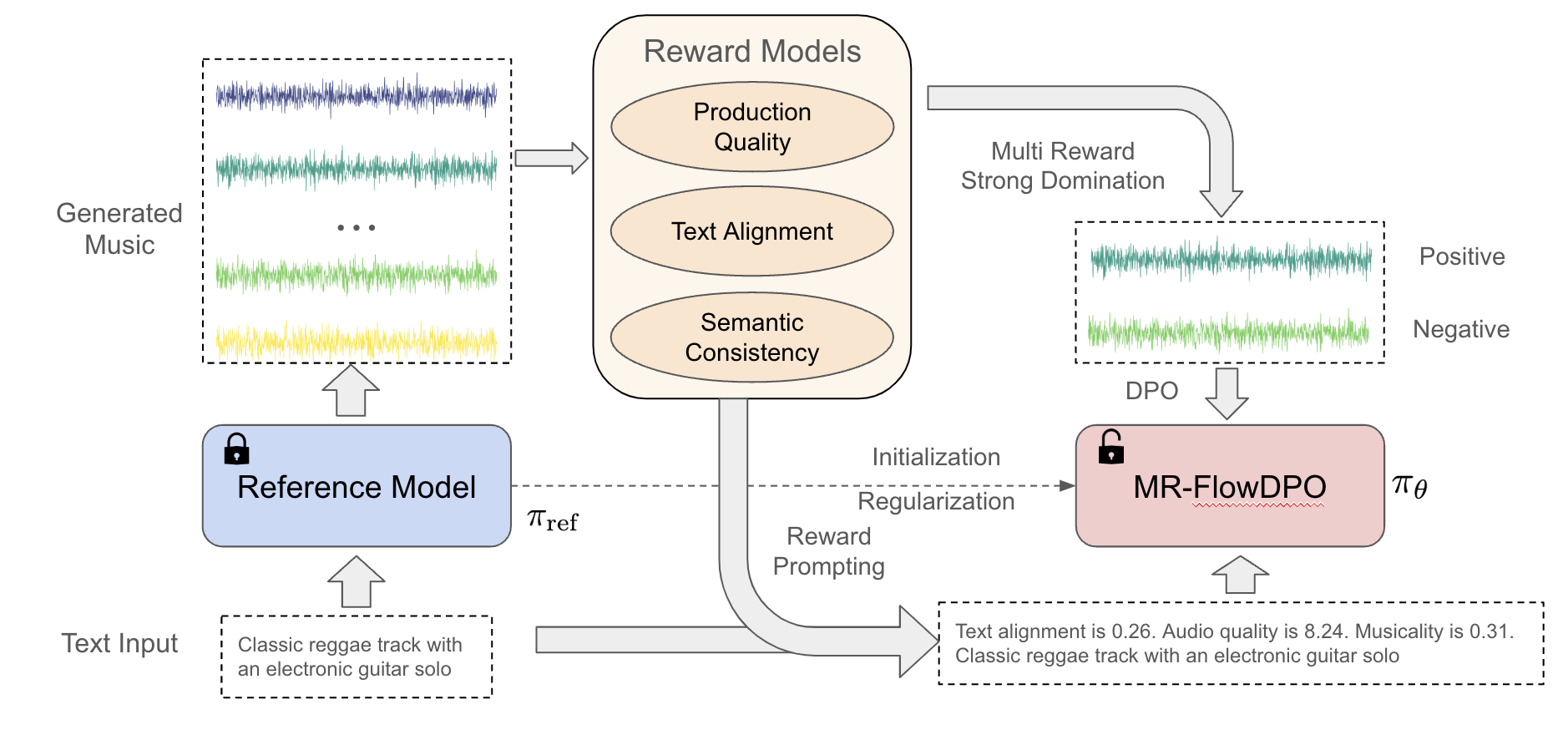}    
    \caption{An overview of MR-FlowDPO. $k$ music samples are generated using the reference model to obtain various automatic reward scores (e.g., production quality), which are later used for preference optimization of \method. \label{fig:arch}}
\end{figure*}

Audio plays a vital role in multimedia content across the internet and daily life, enriching communication, entertainment, and accessibility. It enhances user experiences in applications such as podcasts and immersive sound effects in gaming and film production. With the increasing popularity of generative multimodal AI models (e.g., StableDiffusion~\citep{Esser2024ScalingRF}, SORA~\citep{videoworldsimulators2024}, MovieGen~\citep{Polyak2024MovieGA}), there has been a growing demand for automatic audio generation, both for foley synthesis~\citep{Mei2023FoleygenVA,sheffer2023hear} and as a standalone creative task.  
Among the various sub-modalities of audio, music occupies a central position, serving not only as a form of artistic expression but also as a functional component in applications such as soundtrack composition, podcast production, and adaptive music generation for interactive media.

Advancements in generative modeling, including autoregressive methods~\citep{Brown2020LanguageMA} and diffusion models~\citep{Ho2020Denoising,Song2021scorebased,lipman2023flowmatchinggenerativemodeling}, along with large-scale data collection, have driven rapid progress in audio generation~\citep{agostinelli2023musiclmgeneratingmusictext,copet2024simplecontrollablemusicgeneration,Kreuk2022AudioGenTG,Le2023VoiceboxTM,Vyas2023AudioboxUA,Wang2023NeuralCL,Liu2023AudioLDMTG}.

These advancements have transformed music generation, expanding it from constrained settings (e.g., solo piano generation~\citep{Hawthorne2018EnablingFP}) to open-domain applications~\citep{copet2024simplecontrollablemusicgeneration,agostinelli2023musiclmgeneratingmusictext,liu2023audioldm2,Lan2024HighFT,ziv2024maskedaudiogenerationusing, tal2024jointaudiosymbolicconditioning}, where users can generate fully mixed music based on free-form text descriptions.  
Moreover, models supporting all audio sub-modalities~\citep{liu2023audioldm2,Vyas2023AudioboxUA} (i.e., speech and sound effects) with multimodal prompts~\citep{Polyak2024MovieGA}, have also emerged.

Despite improvements in domain coverage and generation quality, obtaining alignment with human opinion remains a major challenge in music generation.
One of the major limitations stems from the fact that the typical training objectives for such models, does not optimize for human preference, which results in perceptual flaws such as poor text alignment, subpar production quality, and inconsistencies in semantic coherence.

Existing approaches~\citep{cideron2024musicrlaligningmusicgeneration} either rely on human-annotated preference data, which is difficult to scale, or optimize models using a single preference criterion~\citep{majumder2024tango2aligningdiffusionbased,hung2024tangofluxsuperfastfaithful}, limiting their ability to generalize to diverse user preferences.
Guiding models beyond semantic content to other factors at scale - such as recording quality and musical aesthetics - remains largely unexplored.

To address these challenges, we propose \method, a fine-tuning framework for Flow Matching text-to-music generation models.  
Our key contributions are as follows: (1) We propose a multi-reward music preference fine-tuning strategy, integrating multiple perspectives, including text alignment, production quality, and semantic consistency, along with a reward prompting mechanism. In contrast to prior work, which primarily optimizes along \emph{text alignment}, our method optimizes for multiple axes of performance while carefully balancing between them. We conduct a comprehensive ablation study to identify the key parameters in preference data creation, such as margin selection and the balance between different reward dimensions; (2) To address the inherent ambiguity of musicality, which prior work does not quantify directly, we propose a novel self-supervised representation learning method to capture and evaluate semantic consistency in generated music. This leads to significant improvements in rhythmic stability and overall musical coherence; and (3) Applying \method to two strong Flow Matching models, demonstrates consistent and substantial gains across all evaluation dimensions and achieves clear preference over highly competitive baselines.

\begin{figure*}[ht!]
    \centering
    \includegraphics[width=0.9\textwidth]{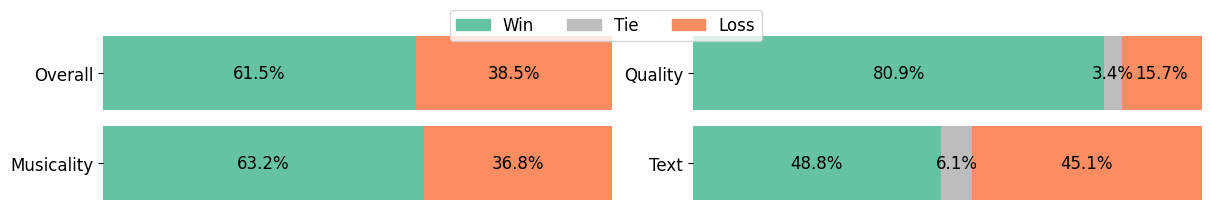}    
    \caption{Human Study - Win Rate of \methMedium against MelodyFlow-1B, evaluated on four axes - Overall preference (Overall), audio quality (Quality), musicality (Musicality) and text alignment (Text).
    \label{fig:bar_plot_against_melodyflow}}
\end{figure*}

\section{Related Work}
\label{sec:prior}

\subsection{Text-to-Music Generation}
Approaches for open-domain text-to-music generation are largely inspired by methods from general audio generative models, 
which can be broadly categorized into auto-regressive audio language modeling and non-autoregressive diffusion modeling.

Auto-regressive audio language models~\citep{agostinelli2023musiclmgeneratingmusictext,Kreuk2022AudioGenTG,copet2024simplecontrollablemusicgeneration,lemercier2024independence} typically utilize a pre-trained audio tokenizer to convert an audio stream into a sequence of discrete tokens. 
These tokens are then processed by a transformer decoder for next-token prediction.  
To better capture music semantics, \citet{agostinelli2023musiclmgeneratingmusictext} has introduced a hierarchical generation framework that decomposes tokens into semantic and acoustic representations, 
employing a cascade of language models for joint training and inference.

Inspired by the success of diffusion models in image generation~\citep{Esser2024ScalingRF}, a large body of research has explored diffusion-based methods for text-to-music generation~\citep{Ho2020Denoising,Song2021scorebased}.  
Noise2Music~\citep{Huang2023Noise2MusicTM} is among the early works adopting this approach, leveraging the Denoising Diffusion Probabilistic Model (DDPM)~\citep{Ho2020Denoising} objective to train a cascaded U-Net that progressively generates audio from lower to higher resolutions.  
Similarly,~\citet{prajwal2024musicflowcascadedflowmatching} proposed a cascaded approach while incorporating flow-matching loss to separately model semantic and acoustic features.  
To improve efficiency, recent work has focused on single-stage high-sample-rate music generation.  
In particular, MelodyFlow~\citep{Lan2024HighFT} and StableAudio~\citep{evans2024stableaudioopen} both utilize VAE-DAC~\citep{Kumar2023highfidelity} to heavily compress high-fidelity music (44.1–48 kHz) into a latent space with low frame rates, enabling the generation of long-duration, high-quality music. 

There have also been works exploring the integration of the above two modeling paradigms. For instance,~\citet{ziv2024maskedaudiogenerationusing} explores a non-autoregressive modeling paradigm over discrete acoustic tokens using a bi-directional transformer encoder.
Additionally,~\citet{Li2023JEN1TU} proposes a multi-task training scheme that alternates between autoregressive and diffusion-based training directions. Recently, both autoregressive and diffusion based approaches have been investigated for fine-grained controlled music generation and editing~\citep{tal2024jointaudiosymbolicconditioning,karchkhadze2024multi,karchkhadze2024simultaneous, rouard2024audio,rouard2025musicgen}.

\subsection{Preference Alignment for Text-to-Audio Generation}
Most existing audio generation models do not incorporate an explicit post-training stage for preference alignment.  
To improve music quality and text faithfulness, prior work has largely relied on the quality of training data.  
For example,~\citet{Vyas2023AudioboxUA,Polyak2024MovieGA} fine-tunes models pre-trained on large-scale datasets using a smaller subset of high-quality data with the same flow matching loss.

An alternative approach applies RLHF to align generative models with human preferences.  
MusicRL~\citep{cideron2024musicrlaligningmusicgeneration} fine-tunes an auto-regressive music language model with human-annotated music preference pairs.
While this method improves performance, it requires a substantial amount of human annotation. In contrast, we leverage \emph{objective} reward metrics to construct fine-tuning data, which are cheaper and easier to scale.  

Similar to our methodology, Tango2~\citep{majumder2024tango2aligningdiffusionbased} and TangoFlux~\citep{hung2024tangofluxsuperfastfaithful} fine-tune pre-trained diffusion or flow-matching models for audio generation.
However, both methods rely solely on text-to-audio alignment score.
In contrast, we optimize along multiple reward dimensions (textual alignment, audio quality, and semantic consistency) and introduce an algorithm to balance between different rewards effectively. 
As will be shown in Section~\ref{sec:ablation_rewards}, incorporating audio quality and semantic consistency rewards is crucial for aligning flow matching music generation.
\section{Method}
\label{sec:method}

\method is based on fine-tuning a text-to-music Flow-Matching \citep{lipman2023flowmatchinggenerativemodeling} model using Direct Preference Optimization (DPO) \citep{rafailov2024directpreferenceoptimizationlanguage}. We construct model-based preference data pairs for text alignment \citep{wu2024largescalecontrastivelanguageaudiopretraining}, audio quality \citep{Tjandra2025A4} and semantic consistency \citep{hsu2021hubertselfsupervisedspeechrepresentation} rewards. The multiple reward scores from the models will be jointly utilized as prompts during inference. In this section, we provide a detailed description of each component.

\subsection{Flow Matching}
Flow matching~\citep{lipman2023flowmatchinggenerativemodeling} is a method for learning the vector field of the transformation between probability densities. Specifically, we follow \citet{lipman2023flowmatchinggenerativemodeling} by considering a Gaussian distribution $p_t(z|z_1)=\mathcal{N}(z|\mu_t(z_1),\sigma_t(z_1)^2I)$ with mean and standard deviation changing linearly in time with $\mu_t(z)=tz$ and $\sigma_t(z)=1-t$, which leads to the target vector field $v_t=z_1 - z_0$, where $z_0\sim \mathcal{N}(0, I)$ and $z_1$ is the target sample. One can learn the vector field by regression using the following
\begin{align}
\label{eq:fm}
L_{\text{FM}}=\mathbb{E}_{t\sim\mathcal{U}(0,1),q(z_1),p(z|z_1)}\|u_t(z;\theta)-v_t\|^2,
\end{align}
where a neural ODE $\frac{d z_t}{dt}=u_t(z_t, \theta)$ is solved to generate samples during inference.

\subsection{DPO for Flow-Matching}
Applying DPO for flow matching consists of using paired data to guide the learning of vector field. Specifically, we adapt the DPO-diffusion~\citep{wallace2023diffusion} loss to the flow matching setup as follows
\begin{align}
      & L_{\text{FM-DPO}}(\theta) = -\mathbb{E}_{t\sim \mathcal{U}(0, 1),x^w, z^l}\log\sigma( \notag \\
      & -\beta(\|u_t(z_t^w;\theta)   -v_t^w\|^2-\|u_t(z_t^l;\theta)-v_t^l\|^2 \notag \\ 
      & -(\|u_t(z_t^w;\theta_{ref})-v_t^w\|^2-\|u_t(z_t^l;\theta_{ref})-v_t^l\|^2))), \notag \\ 
\end{align}
where $t$ is the flow matching timestamp, $(z^w, z^l)$ are positive and negative samples respectively, and $(v^w, v^l)$ are the positive and negative target vector fields respectively. 

Compared to~\citet{wallace2023diffusion}, we calculate the difference in vector fields instead of the difference in noise, similarly to~\citet{hung2024tangofluxsuperfastfaithful}. 

\subsection{Reward functions}
In this section, we focus on the construction of paired data, which involves designing multi-dimensional scoring functions aimed at: (i) text alignment; (ii) audio quality; and (iii) semantic coherence for data pairs and developing a reward combination algorithm for generating these pairs.

\subsubsection{Text reward}
Given an audio-text pair, we assign the text reward to be the cosine similarity between the audio and text embeddings, extracted by the CLAP model~\citep{wu2024largescalecontrastivelanguageaudiopretraining}. Specifically, this score measures how a generated audio aligns to the text description. Text reward has been widely used in prior works~\citep{cideron2024musicrlaligningmusicgeneration, majumder2024tango2aligningdiffusionbased, hung2024tangofluxsuperfastfaithful} for text-to-audio generation. Specifically for music, we aim at maximizing the alignment on the musical elements (e.g., genre) specified by the text, thus we employ the CLAP model~\footnote{\url{https://huggingface.co/lukewys/laion_clap/blob/main/music_audioset_epoch_15_esc_90.14.pt}} trained on music data, where in Section~\ref{sec:results}, we will validate such choice.

\subsubsection{Production quality reward}
Production quality refers to the acoustic characteristics of music, distinct from subjective quality. 
High-quality audio should have clear, well-defined elements with minimal noise, distortion, or artifacts.

To measure production quality, we use the aesthetic score predictor proposed by~\citet{Tjandra2025A4}, a regression transformer trained on $500$ hours of general audio from various domains, including music. 
The training data annotation for the aesthetic predictor considers multiple acoustic dimensions, including clarity, dynamics, frequency balance, spatialization, and overall technical proficiency. 
The model outputs a scalar score between $1$ and $10$, where higher values indicate better production quality. As shown by~\citet{Tjandra2025A4}, this predictor is highly correlated with human judgments of production quality.

\subsubsection{Semantic Consistency Reward}
Preliminary experiments show that generated samples often lack semantic and rhythmic consistency (see Section~\ref{sec:results}). We hypothesize this occurs as \method relies on flow-matching without conditioning on semantic representation. To address this, we introduce a novel \emph{Semantic Consistency Reward} function to enhance musicality without requiring multi-stage cascade modeling.

We construct preference pairs defined by the likelihood of a HuBERT \citep{hsu2021hubertselfsupervisedspeechrepresentation} model.
HuBERT learns a semantic, contextualized audio representation, using a self-supervised masked language modeling objective. Semantic representations such as HuBERT have been shown to be effective in maintaining structural coherence in speech and music generation \citep{lakhotia2021generativespokenlanguagemodeling, borsos2023audiolmlanguagemodelingapproach, agostinelli2023musiclmgeneratingmusictext, zeldes2024enhancing, prajwal2024musicflowcascadedflowmatching}.
Unlike prior work, we improve semantic coherence through DPO supervision, without the need for cascaded modeling. 

As the original HuBERT model~\citep{hsu2021hubertselfsupervisedspeechrepresentation} is trained on speech, we re-train the model on music data to enhance its music understanding capacity following the same pipeline. We train HuBERT on the training set used for the reference model. We stick to the 24-layer HuBERT model configuration ("Large" scale as defined in \citet{hsu2021hubertselfsupervisedspeechrepresentation}). We cluster the 12th layer of the trained model into 1024 centroids defined by $e_1, ..., e_{1024}$ using the k-means algorithm \citep{Hartigan1979}.
Given an audio $X$, we first apply an unmasked forward pass of the HuBERT model. We then compute the cosine similarity between the $n$th-layer features, defined by $o_1, ..., o_T$ and the $n$th-layer centroids, and extract a per-centroid probability defined by:
\begin{equation}
    p(c \mid X, t) = \frac{\exp(\text{sim}(o_t, e_c) / \tau)} {\sum_{c'=1}^{N} \exp(\text{sim}(o_t, e_{c'}) / \tau)},
\end{equation}
with $\tau$ being the temperature and $N$ being the number of centroids.
The \emph{semantic consistency score} $\bar{p}$ is defined as:
\begin{equation}
    \bar{p} := \frac{1}{T} \sum_{t=1}^{T}max_{c}( \log p(c \mid X, t)).
\end{equation}

Intuitively, this reward measures the semantic coherence of the generated music and consists of two steps:
(1) extracting the most likely sequence of music tokens, and (2) scoring the likelihood of that sequence.

For (1), discrete tokens from HuBERT trained on music data has shown to capture musical semantics such as melody and rhythm~\citep{agostinelli2023musiclmgeneratingmusictext, prajwal2024musicflowcascadedflowmatching}.
For (2), we treat the HuBERT model as an implicit language model, which is trained with a masked language modeling objective and encodes a strong prior over plausible musical sequences. The token sequence likelihood under this model serves as a proxy for semantic coherence in the generated music.

\subsection{Preference Data Creation}
\label{subsec:dpo_data_creation}
\textbf{Reference model sampling and reward evaluation.}
Given a set of $N$ text prompts sampled uniformly at random from the training set, we generate $k$ samples per prompt with the reference model, to create a total samples pool of $k \cdot N$ samples, for which we evaluate each of the reward axes.  

\textbf{Extracting per-reward margin thresholds.} Given a text prompt, there are at most $\binom{k}{2}$ pair candidates. For each reward axis $r$, we calculate the set of all $N \cdot \binom{k}{2}$ difference values, and extract a non-negative threshold $\tau_r$, defined as the $95$-percentile over the set of differences in absolute value.
To prevent using outlier negatives, such as noise and silence, we define a minimal threshold for each reward, defined by the $5$-percentile of the reward scores. For the semantic consistency reward, we additionally apply maximum thresholding. We noticed that high values of the HuBERT-based reward correlate with silence as well as noise. 

\textbf{Multi-reward pairing with strongly dominating positives.}
We propose a multi-reward pairing scheme that guarantees domination of the positive sample over the negative one in all reward axes (Text alignment, audio quality and semantic consistency), inspired by the methodology suggested in Parrot \citep{lee2024parrotparetooptimalmultirewardreinforcement}. In our pairing methodology, which we call Multi Reward Strong Domination (MRSD) (See Algorithm~\ref{alg:msrd}), we extract pairs for which the margin in at least one axis of reward $s$, which we call the primary axis, is larger than the primary threshold $\tau_s^p$, which we set as the 95-percentile of the score difference. Additionally, we ensure that the positive sample reward exceeds the negative sample reward across all secondary axes $s^\prime$ with a smaller margin. This threshold $\tau_{s^\prime}^n$ is defined by the median absolute reward difference.
Finally, we down-sample the resulting pairs to balance the primary axes. The resulting dataset consists of $3\cdot R$  triplets $(X^{w}, X^{l}, Y)$ , where $X^{w}$ is the strongly dominating positive samples, $X^{l}$ is the negative samples, and $Y$ is the textual description. Each of the three reward axes - text-aligment, audio quality and semantic consistency - contributes $R$ triplets as the primary axis of domination. 

\begin{algorithm}[t!]
\caption{Multi Reward Strong Domination}
\label{alg:msrd}
\begin{algorithmic}[1] 
    
    \Require a set of text prompts $\mathcal{T}$, reference model $\mathcal{M}$, \\
    principle thresholds $\pmb\tau^p$, secondary thresholds $\pmb\tau^n$
    
    \Ensure Strong Dominant set $\mathcal{P}$
    
    \State $\mathcal{P} \gets \emptyset$
    \State $S \gets \{\text{\small semantic consistency}, \text{audio quality}, \text{text alignment}\}$
    
    \For{$y\in \mathcal{T}$}
        \State Generate $k$ samples: $\{{\bf x}^1,...,{\bf x}^k\} := \mathcal{M}(y)$
        \For{$s\in S$}
            \State $\mathcal{P}_s\gets \emptyset$
            \For{$i=1$ to $k$, $j=1$ to $k$}
                \If{$|r_s(x^i)-r_s(x^j)|>\tau^p_s$ \textbf{and} $|r_{s^\prime}(x^i)-r_{s^\prime}(x^j)|>\tau^n_{s^\prime}, \forall s^\prime\in S \backslash \{s\}$}
                    \State $\mathcal{P}_s \gets \mathcal{P}_s\cup\{(x^i, x^j)\}$
                \EndIf
            \EndFor
            \State Sample $R$ samples from $\mathcal{P}_s$ into $\mathcal{P}$
        \EndFor
    \EndFor

\end{algorithmic}
\end{algorithm}

\subsection{Reward Prompting}
We prepend the original textual description with a predefined prompt, in natural language, describing the different reward values. An example of such prompt would be \textit{"Text alignment is 0.26, Audio quality is 8.24, Semantic Consistency is 0.31"}.  During DPO, we use the reward values of the positive audio sample, while at inference time we use the 99-percentile taken over these values, corresponding to exceptionally high rewards.  

\section{Experimental Setup}
\label{sec:exp_setup}

\subsection{Setup}
\textbf{Reference Models} We experiment with DPO of two reference models: (i) a flow-matching model (Flow-400M) of $400$M parameters trained from scratch on large-scale music data,
and (ii) the pre-trained MelodyFlow-1B ("MelodyFlow-Medium" in \citep{lan2024highfidelitytextguidedmusic}), one of the SOTA flow-matching-based music generation models. Training details of these two models are given in the Appendix.

\textbf{Pre-training data}
We follow a similar experimental setup as in \citep{copet2024simplecontrollablemusicgeneration, ziv2024maskedaudiogenerationusing, tal2024jointaudiosymbolicconditioning}, and use a training dataset consisting of $20$K hours of licensed music from the Shutterstock and Pond5 \footnote{\url{shutterstock.com/music} and \url{pond5.com}} with $25$K and $375$K instrument-only music tracks, respectively. All datasets are sampled at $32$kHz, paired with textual descriptions. 

\textbf{DPO data} We construct preference datasets according to the MRSD pairing methodology presented in Section \ref{subsec:dpo_data_creation}, with CLAP, Production Quality Aesthetics and HuBERT-likelihood rewards, $N=20K$ text descriptions randomly sampled from the training set, $k=16$ and with $R=30K$ samples per primary axis of reward. Overall, each model-based preference dataset consists of 90K triplets of the form $(X^{w}, X^{l}, Y)$.  

\textbf{Evaluation data} For evaluation, we use the MusicCaps benchmark \citep{agostinelli2023musiclmgeneratingmusictext}, comprised of $5.5$K $10$-second samples. For objective evaluation we use the entire dataset, while for human evaluation we curated 100 samples using the $1$K genre-balanced following~\citep{agostinelli2023musiclmgeneratingmusictext}, while removing text prompts that mention vocals, low quality audios. Additionally, we use a held-out test set of $1$K instrumental tracks from a proprietary high-quality data, on which we perform our ablation studies. 

\textbf{DPO fine-tuning}
We fine-tune each reference model for $10$ DPO epochs with $\beta=2000$. We use a learning rate peak of $1e-6$, with a linear warmup of $1K$ training steps, and a linear decay for the remaining part of the DPO. Following prior work~\citep{majumder2024tango2aligningdiffusionbased}, we use a batch size of 32, and an AdamW optimizer with $(\beta_1, \beta_2)= (0.9,0.999)$, $\epsilon=1e-8$ and weights decay $1e-2$.

\vspace{-0.1cm}
\begin{table}
\caption{Objective Evaluation: Comparison of \method to prior work and baselines, reported on MusicCaps.}
\label{objective_eval_table}
\centering
\resizebox{0.7\textwidth}{!}{
\begin{tabular}{lcccccr}
\toprule
Method & Aes $\uparrow$ & EA $\uparrow$ & CLAP $\uparrow$ & BPM-std $\downarrow$ & FAD $\downarrow$ \\
\midrule
MusicGen   & 7.17 & 6.72 & 0.29 & 7.60 & \textbf{4.69} \\
MelodyFlow-1B & 7.13 & 6.69 & 0.29 & 8.01 & 4.96 \\
AudioLDM2  & 7.10 & 5.88 & \textbf{0.30} & 7.66 & 5.14 \\
\midrule
Flow-400M        & 7.08 & 6.50 & 0.29 & 9.09 & 2.70 \\
Flow-400M+RP     & 8.25 & 7.08 & 0.27 & 8.67 & 8.73 \\
OnlySFT    & 6.91 & 6.13	& 0.30	& 10.14 & 3.38 \\
\methSmall & 8.10 & 7.18 & 0.28 & 7.57 & 6.47 \\
\methMedium & \textbf{8.26} &	\textbf{7.72}	& 0.27	& \textbf{6.11} & 11.26 \\
\bottomrule
\end{tabular}
}
\end{table}

\subsection{Objective Metrics}
\textbf{Aesthetic Scores.}
We use their production quality score as a metric for audio quality. To evaluate semantic consistency, we adopt the associated Content Enjoyment Aesthetics (EA) score, which is trained directly on human ratings of the subjective listening experience. \citet{Tjandra2025A4} also showed that the two scores correlate highly with human judgments.

\textbf{BPM-std.} We propose using the variance of a Beats Per Minutes (BPM) estimation, when taken over different local regions of the audio waveform, as a metric for rhythmic stability. We use a BPM estimator implementation from the librosa \citep{mcfee2015librosa} package. Given an audio sample, we extract the BPM of every non-overlapping $3.33$ seconds window, and compute the standard deviation over the different local estimations. 

\textbf{Text Alignment and FAD.} We measure text-adherence using the CLAP score as defined in Section \ref{sec:method}. Additionally, we report Fréchet Audio Distance (FAD) \citep{kilgour2019frechetaudiodistancemetric}, a perceptual metric for audio quality commonly used in prior audio generation research.   

\vspace{-0.1cm}
\subsection{Human Evaluation}
We perform subjective evaluation of relative preference measuring four aspects of performance: Overall preference, audio quality, text alignment and musicality. We use professional human annotation service. For each configuration, we evaluate $100$ triplets of the form $(X^{A}, X^{B}, Y)$.  We collect $5$ annotations per pair, per question resulting in a total of $500$ responses to $4$ questions. For evaluating winning rate, we apply cross-annotator consensus analysis, and for statistical scoring we use soft scores and extract the mean and $95$ confidence interval of the soft random variable representing the winning model. Full details on the annotation process can be found in appendix.

\vspace{-0.1cm}

\section{Results}
\label{sec:results}

\subsection{Comparison to Baselines}
We compare $\method$ to MusicGen \citep{copet2024simplecontrollablemusicgeneration}
, MelodyFlow ~\citep{lan2024highfidelitytextguidedmusic}
, AudioLDM2 ~\citep{liu2023audioldm2}
. In addition, we compare 
\method to (i) a reference model, pretrained with the proposed reward prompting strategy (Flow-400M+RP), and to (ii) a baseline in which we replace DPO with positive samples only fine-tuning of the reference model (OnlySFT). In Table \ref{objective_eval_table}, we report objective metrics on MusicCaps \citep{agostinelli2023musiclmgeneratingmusictext}.  Results suggest that \method is better aligned with the perceptual audio production quality (Aes) and content enjoyness (EA) rewards, while performing slightly worse than several baselines in terms of text-aligment. Additionally, it can be seen that \method effectively mitigates the rhythmic instability of the reference Flow-Matching model, reducing the BPM-std metric to a value which is on par or better than all baselines.

\subsection{Human Evaluation}
The main results of the human evaluation are presented in Table \ref{tab:merged_winrate}, in which we compare \methSmall to prior work and various baseline models.

We measure the net win rate (win\% - loss\% of our model) which has a range of $[-100\%,100\%]$, following~\citep{Polyak2024MovieGA}.
Specifically, for each item, we take the mean of preference between two models  as the consensus score. We bootstrap item-level consensus score $1000$ times, where the mean consensus score is taken and $95\%$ confidence interval is the difference between $2.5\%$-ile and $97.5\%$-ile of the net win rate. 

Results suggest that \methSmall performs significantly better than the reference Flow-400M model in all four axes (Overall preference, audio quality, text alignment and musicality), better than MusicGen-medium in audio quality (on-par in the other axes), and significantly better than MelodyFlow-1B and AudioLDM2 in all aspects.

Additionally, we compare \methMedium to its reference model - MelodyFlow-1B. The overall trends remains generally consistent with \methSmall, where we achieve statistically significant gains in overall generation quality, musicality and production quality, while being comparable in text alignment. The lack of improvement in text alignment is expected, as the reference model MelodyFlow-1B already performs well in this aspect.

\begin{table}
    \caption{Net win rate (\%) of \methSmall and \methMedium against baselines.}
    \label{tab:merged_winrate}
    \centering
    \resizebox{.9\textwidth}{!}{
    \begin{tabular}{lccccc}
        \toprule
        & \methSmall & \multicolumn{3}{c}{\methSmall} & \methMedium  \\
        \cmidrule(r){2-2} \cmidrule(r){3-5} \cmidrule(r){6-6}
        & Flow-400M & MusicGen & MelodyFlow & AudioLDM2 & MelodyFlow \\
        \midrule
        Overall Preference & 25.02\pmrtiny{12.00} & 2.23\pmrtiny{11.70} & 36.67\pmrtiny{10.70} & 41.08\pmrtiny{10.60} & 16.67\pmrtiny{10.00} \\
        \midrule
        Audio Quality      & 12.46\pmrtiny{12.40} & 17.09\pmrtiny{10.30} & 56.72\pmrtiny{7.70}  & 71.02\pmrtiny{6.40}  & 43.26\pmrtiny{10.50} \\
        \midrule
        Text Alignment     & 24.10\pmrtiny{11.60} & -2.88\pmrtiny{10.20} & 15.04\pmrtiny{10.60} & 12.62\pmrtiny{10.00} & 1.88\pmrtiny{9.30}   \\
        \midrule
        Musicality         & 20.37\pmrtiny{12.30} & 2.65\pmrtiny{11.30}  & 32.66\pmrtiny{11.20} & 38.24\pmrtiny{10.50} & 17.00\pmrtiny{10.30} \\
        \bottomrule
    \end{tabular}
    }
    \vspace{-0.2cm}
\end{table}

\subsection{Ablation Study and Analysis}

\subsubsection{Ablation on rewards}
\label{sec:ablation_rewards}
In Table~\ref{tab:multi_reward_and_margin}, we compare different multi-reward configurations. We fine-tune the reference model with DPO with (i) Text-reward only (+TR), (ii) Text and audio production quality rewards (+TR+AR) and (iii) \method, including all three reward axes. 

\textbf{TR only hurts rhythmic stability}
Performing DPO with TR leads to an improvement of the reference model in CLAP score, and a slight improvement in the two Aesthetic scores. On the other hand, such an optimization leads to a degradation in rhythmic stability (BPM-std: $7.77\rightarrow 8.40$)

\textbf{TR+AR+SR achieves the best result}
While Ref+TR achieves the highest CLAP score, and Ref+TR+AR achieves the highest Aesthetics score, both configurations struggle with rhythmic stability, indicated with a high BPM-std. \method obtains a significantly lower BPM-std score, while maintaining an on-par or better Aesthetic and CLAP scores.

\begin{table}
\caption{Effect of incorporating different rewards (left) and reward margin ablation (right), evaluated on the in-domain test set.}
\label{tab:multi_reward_and_margin}
\centering
\scriptsize
\begin{minipage}{0.45\textwidth}
\centering
\setlength{\tabcolsep}{4pt}
\begin{tabular}{lccccc}
\toprule
Method & Aes $\uparrow$ & EA $\uparrow$ & CLAP $\uparrow$ & BPM $\downarrow$ & FAD $\downarrow$ \\
\midrule
Ref & 7.58 & 6.90 & 0.33 & 7.77 & \textbf{0.78} \\
+ TR & 7.80 & 7.21 & \textbf{0.38} & 8.40 & 1.04 \\
+ TR + AR & \textbf{8.33} & 7.47 & 0.35 & 8.06 & 1.99 \\
+ TR + AR + SR & 8.26 & \textbf{7.55} & 0.37 & \textbf{6.00} & 1.76 \\
\bottomrule
\end{tabular}
\end{minipage}
\hfill
\begin{minipage}{0.45\textwidth}
\centering
\setlength{\tabcolsep}{4pt}
\begin{tabular}{lccccc}
\toprule
Perc & Aes $\uparrow$ & EA $\uparrow$ & CLAP $\uparrow$ & BPM $\downarrow$ & FAD $\downarrow$ \\
\midrule
Ref & 7.58 & 6.90 & 0.33 & 7.77 & \textbf{0.78} \\
25 & 7.95 & 7.32 & 0.36 & \textbf{6.55} & 1.27 \\
50 & 7.99 & 7.27 & 0.36 & 6.95 & 1.52 \\
75 & 8.03 & 7.21 & 0.36 & 7.14 & 1.16 \\
95 & \textbf{8.07} & \textbf{7.42} & \textbf{0.38} & 7.45 & 1.22 \\
\bottomrule
\end{tabular}
\end{minipage}
\vspace{-0.2cm}
\end{table}

\subsubsection{Reward Margin.}
In Table \ref{tab:multi_reward_and_margin} we present an ablation study on the reward difference used during the model-based preference data creation of $\method$. We report results on the in-domain test-set comparing reward difference margins of percentiles $25, 50, 75$ and 95. We create the data without applying the \paretomethod criteria. We run each configuration for $20$ DPO epochs.

Results suggest that using larger reward margin leads to a stronger improvement in Aesthetics, EA and CLAP scores. The trend in BPM-std is the opposite: The larger the margin, the worse the rhythmic stability measured by BPM-std. For reward margin percentile $25$, the BPM-std of the reference model was reduced in $\sim 15\%$, while for reward margin $95$ the reduction was only $\sim4\%$. We hypothesize that without \paretomethod criteria, the Clap and Aesthetics supervision dominates the DPO process, while the influence of the semantic consistency reward is milder. 

\subsubsection{Using Multi Reward Strongly Dominating Positives}
In Table~\ref{tab:prompting_pareto} we present an ablation study comparing $\method$ to $\method_{\text{w/o \paretomethod}}$ - where in the latter we do not apply the secondary axes domination criterion during the preference data creation.

Results suggest that with \paretomethod, the multi-reward preference optimization is faster, and the different reward axes are better balanced. Specifically,  with \paretomethod, there is a joint improvement in inherently different axes of performance - (i) Audio quality, represented by the Aesthetic scores, is increased from $7.99$ to $8.26$,  and (ii) rhythmic stability, represented by the semantic HuBERT reward, and measured by a reduced BPM-std of $6.00$ compared to $6.84$  of $\method_{\text{w/o \paretomethod}}$. 

\begin{table}
\caption{Effect of reward prompting (left) and \paretomethod{} criteria (right), reported on the in-domain test set.}
\label{tab:prompting_pareto}
\centering
\resizebox{\textwidth}{!}{
\begin{tabular}{cc}
\begin{tabular}{lccccc}
\toprule
Method & Aes $\uparrow$ & EA $\uparrow$ & CLAP $\uparrow$ & BPM $\downarrow$ & FAD $\downarrow$ \\
\midrule
Ref & 7.58 & 6.90 & 0.33 & 7.77 & \textbf{0.78} \\
\textsc{MR-FlowDPO}\textsubscript{w/o Prompting} & 8.20 & 7.21 & 0.35 & 7.30 & 2.10 \\
\textsc{MR-FlowDPO} & \textbf{8.26} & \textbf{7.55} & \textbf{0.37} & \textbf{6.00} & 1.76 \\
\bottomrule
\end{tabular}
&
\begin{tabular}{lccccc}
\toprule
Method & Aes $\uparrow$ & EA $\uparrow$ & CLAP $\uparrow$ & BPM $\downarrow$ & FAD $\downarrow$ \\
\midrule
Ref & 7.58 & 6.90 & 0.33 & 7.77 & \textbf{0.78} \\
\textsc{MR-FlowDPO}\textsubscript{w/o \paretomethod} & 7.99 & 7.43 & 0.37 & 6.84 & 0.76 \\
\textsc{MR-FlowDPO} & \textbf{8.26} & \textbf{7.55} & \textbf{0.37} & \textbf{6.00} & 1.76 \\
\bottomrule
\end{tabular}
\end{tabular}
}
\end{table}

\subsubsection{Effect of prompting}
In Table \ref{tab:prompting_pareto} we present an ablation study comparing $\method$ to $\method_{\text{noRP}}$ - where in the latter the reward prompting mechanism is omitted from both training and inference. Results show a clear improvement in all metrics, obtained by applying our reward prompting mechanism.

\section{Conclusion}
\label{sec:results}

In this paper, we propose \method, a fine-tuning method for text-to-music flow matching models incorporating multiple rewards. 
\method utilizes three key reward modules to evaluate generated audio: a CLAP model for text alignment, a production quality predictor for assessing audio fidelity, and a HuBERT-based semantic module for measuring semantic consistency. To effectively integrate these scores, we introduce a novel pairing scheme, MRSD, which retains only dominant pairs across all three axes. Extensive experiments highlight the importance of data filtering and MRSD pairing in fine-tuning. We demonstrate that \method significantly improves overall music generation quality and is consistently preferred over strong baselines in terms of audio quality, text alignment, and semantic consistency.
\section*{Impact Statement}
\label{sec:impact}
The ability to generate high-quality, coherent music from text descriptions has numerous positive applications, including content creation, accessibility, and personalized music experiences. However, as with any generative model, ethical considerations must be taken into account.
We emphasize the importance of using licensed datasets to mitigate the risks on fair use, licensing, and ownership of generated content. Additionally, generative models may inadvertently reinforce biases present in training data, such as genre imbalances or cultural under-representation. Future work should focus on improving dataset diversity and fairness in music generation.

\bibliographystyle{unsrtnat}
\bibliography{refs}

\newpage

\appendix
\clearpage
\section{Experimental setup}
\label{app:sec:exp}
\subsection{Subjective Evaluation}
\begin{figure}[htp]
    \centering
    \includegraphics[width=\linewidth]{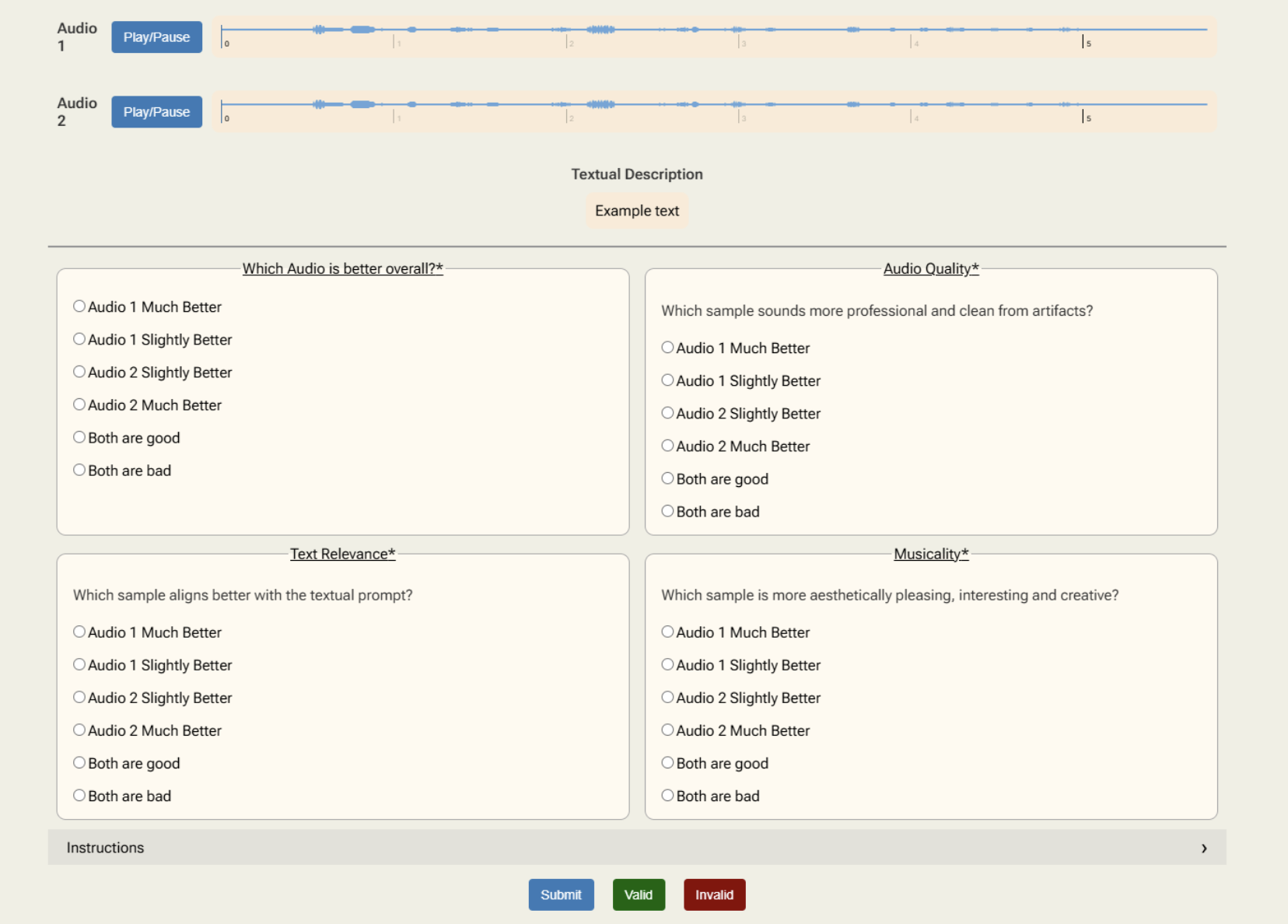}
    \caption{Human Annotation User Interface}
    \label{fig:annotation_ui}
\end{figure}

Below is the guideline for subjective evaluation (see Figure~\ref{fig:annotation_ui} for UI)

\begin{itshape}
For each task, you will listen to a pair of audio clips, each 10-30 seconds long. 
The audio content is instrumental music of a variety of genres. 
You would also observe a textual description used for generating the two samples.
Your task is to:
\begin{itemize}
    \item Determine your overall relative preference between the samples.
    \item Determine your relative preference in terms of each of the following aspects:
    \begin{itemize} 
        \item Audio Quality - Which sample sounds more professional and clean from artifacts?  
\item Text Relevance - Which sample aligns better with the textual prompt?
\item Musicality - Which sample is more aesthtically pleasing, interesting and creative?  
    \end{itemize}
\end{itemize}

\end{itshape}

\subsubsection{Further Details on Win-Rate analysis}
In the main paper we present results of "Human Study - Win Rate of \methMedium against MelodyFlow-1B". For measuring the win-rate, we apply an infra-annotator consensus analysis between the 5 annotators, to filter out ambiguous pairs. Among the non-ambiguous pairs, for each performance axis, we report "Win" of \methMedium if the most frequent preference was to \method, "Tie" if the leading answer was "both are good" or "both are bad", and "Loss" if the most frequent preference was to MelodyFlow-1B. Results suggest that \methMedium is superior to the reference model MelodyFlow-1B, in all performance axes.

\section{Results}

\subsection{In-domain test set comparison}

In Table \ref{objective_eval_indomain_table} we further compare \methMedium to its reference model MelodyFlow-1B, the corresponding reference models on the in-domain test set.  
In both models, applying \method yields similar gains as to the Flow-400M, which suggests the effectiveness of \method regardless of the reference model.

\begin{table}
\caption{Objective Evaluation: Comparison of \method to reference models, reported on the in-domain test set. }
\label{objective_eval_indomain_table}

\centering
\resizebox{0.7\textwidth}{!}{
\begin{tabular}{lcccccr}
\toprule

Method & Aes $\uparrow$ & EA $\uparrow$ & CLAP $\uparrow$ & BPM-std $\downarrow$ & FAD $\downarrow$ \\

\midrule
Flow-400M     & 7.58 & 6.90 & 0.33 & 7.77 & \textbf{0.78} \\
MelodyFlow-1B & 7.34 & 7.08 & 0.35 & 7.54 & 1.73 \\
\midrule
\methSmall    & 8.26 & 7.55 & \textbf{0.37} & 6.00 & 1.76 \\
\methMedium   & \textbf{8.29} & \textbf{7.82} & \textbf{0.37} & \textbf{5.79} & 2.73 \\
\bottomrule
\end{tabular}
}
\end{table}

\subsection{Music reward comparison}
We compare \method against the alternative of using the Content Enjoyment Aesthetics (EA) score for musicality. In Table \ref{musicality_hubert_vs_enjoyness} we present results on the in-domain test-set comparing $\method$ to $\method_{\text{EA}}$ - where in the latter we replace the HuBERT-based semantic consistency rewards with the EA scores. Results suggest that using EA as a musicality reward leads to over-optimization of audio quality, while degrading the BPM-std of the resulting music, from $7.77$ to $8.19$. In contrast, the proposed HuBERT-based semantic consistency reward acts as a complementary axis of reward, enabling enhancement of rhythmic stability. 

\begin{table}
\caption{Comparison between using Music-HuBERT and Enjoyness Aesthetic for musicality reward}
\label{musicality_hubert_vs_enjoyness}
\centering
\resizebox{0.7\textwidth}{!}{
\begin{tabular}{lccccr}
\toprule
Method & Aes $\uparrow$ & EA $\uparrow$& CLAP $\uparrow$& BPM-std $\downarrow$ & FAD $\downarrow$ \\
\midrule
Ref & 7.58& 6.90& 0.33& 7.77& \textbf{0.78}\\
\method\textsubscript{EA}& \textbf{8.39}& \textbf{7.70} & \textbf{0.38}& 8.19& 2.32\\
\method & 8.26& 7.55& 0.37& \textbf{6.00}& 1.76\\
\bottomrule
\end{tabular}
}
\end{table}

\subsection{Ablation Study and Analysis}
\subsubsection{Effect of regularization with Flow-Matching loss}
The authors of TangoFlux [see reference in the main paper] suggest adding the flow-matching loss term of the positive sample as a regularization to DPO. In a preliminary experiment, we tested the performance of MR-FlowDPO when such an additive term is added to the loss function. In Table~\ref{tab:objective_eval_fm_loss_reg}, we compare \methSmall to \methSmall with such a regularization term (\methSmall w. FM-loss regularization). Results suggest that the regularization term does not lead to a significant improvement in CLAP as opposed to the trend reported in TangoFlux. 

\begin{table}
\caption{Objective Evaluation: Effect of regularization with Flow-Matching loss. }
\label{tab:objective_eval_fm_loss_reg}
\centering
\resizebox{0.8\textwidth}{!}{
\begin{tabular}{lcccccr}
\toprule

Method & Aes $\uparrow$ & EA $\uparrow$ & CLAP $\uparrow$ & FAD $\downarrow$ \\

\midrule
Flow-400M                              & 7.58 & 6.90 & 0.33 & 0.78 \\
\methSmall w. FM-loss regularization & 8.10 & 7.16 & 0.37 & 1.36 \\
\methSmall                             & 8.08 & 7.31 & 0.38 & 1.23 \\
\bottomrule
\end{tabular}
}
\end{table}

\subsubsection{Effect of Continuous Values in Reward Prompting}
In Table~\ref{tab:objective_eval_reward_prompt_binary_vs_continuous_reg}, we present results of a preliminary experiment we conducted testing an alternative representation for the reward textual prompting. As opposed to the approach presented in the methodology section, in which we prepend continuous reward values to the textual prompt, we report results on a binarized approach, in which we add textual prompts such as "Text alignment is better", "Audio quality is worse" and "Musicality is much better". Results suggest that the proposed continuous reward textual prompting approach is more effective than the binarized alternative. 

\begin{table}
\caption{Objective Evaluation: Effect of Continuous Values in Reward Prompting. }
\label{tab:objective_eval_reward_prompt_binary_vs_continuous_reg}
\centering
\resizebox{0.8\textwidth}{!}{
\begin{tabular}{lcccccr}
\toprule

Method & Aes $\uparrow$ & EA $\uparrow$ & CLAP $\uparrow$ & FAD $\downarrow$ \\

\midrule
Flow-400M                    & 7.58 & 6.90 & 0.33 & 0.78 \\
\methSmall w. Binary rewards & 8.07 & 7.19 & 0.37 & 1.70 \\
\methSmall                   & 8.25 & 7.55 & 0.37 & 1.76 \\
\bottomrule
\end{tabular}
}
\end{table}

\subsection{Semantic consistency reward construction}
We hereby add results from a preliminary experiment we conducted, comparing different statistics over the semantic representation for musicality reward (or semantic consistency reward). We implemented four scoring functions derived from HuBERT, all use the same feature representation described in the methodology section: (i) Local span masking - in which we measure the likelihood of masked spans given unmasked spans, where the span length is the minimal used for HuBERT training, (ii) Global span masking - scoring how likely is one half of the audio given the other half, (iii) Random span masking - in which we use the exact same masking scheme used for HuBERT training and score the likelihood of masked timesteps given the unmasked ones, and finally (iv) which is the approach described in the methodology section ("unmasked HuBERT"). In Table~\ref{tab:objective_eval_semantic_score_function}, we compare the reference model Flow-400M, to \methSmall with each of the four semantic consistency scoring functions. Results suggest that the unmasked statistic is the most efficient in improving production quality, content enjoyment and text alignment of the reference model. 

\begin{table}
\caption{Semantic Consistency Reward - ablation on score function}
\label{tab:objective_eval_semantic_score_function}
\centering
\resizebox{0.8\textwidth}{!}{
\begin{tabular}{lcccccr}
\toprule

Method & Aes $\uparrow$ & EA $\uparrow$ & CLAP $\uparrow$ & FAD $\downarrow$ \\

\midrule
Flow-400M                         & 7.58 & 6.90 & 0.34 & 0.78 \\
\methSmall w. Local span masking  & 7.98 & 7.16 & 0.37 & 1.15 \\
\methSmall w. Global span masking & 8.00 & 7.05 & 0.37 & 1.16 \\
\methSmall w. Random span masking & 7.94 & 7.27 & 0.37 & 1.04 \\
\methSmall (unmasked HuBERT)      & 8.04 & 7.33 & 0.38 & 1.10 \\
\bottomrule
\end{tabular}
}
\end{table}

\subsection{Comparing MRSD to Human annotated pairs.}
We compare \method's pairing methodology (MRSD), to the alternative of direct human annotation. For the alternative approach, we collected 32K pairs annotated by humans. In Table~\ref{tab:mrsd_vs_supervised_pairs} we report results on (i) DPO with the annotated data, (ii) \methSmall when data is created by applying MRSD to the exact same samples pool used in the annotated data ("\methSmall w. 32K MRSD pairs"), and (iii) \methSmall with a 90K pairs created from a larger samples pool using the MRSD algorithm. We observed no performance improvements obtained by running DPO on the 32K pairs we collected, compared to the proposed alternative of multi-reward model based pairing via the MRSD algorithm. Even under the same scale (32K), using proposed method leads to similar performance as using human annotated data.

\begin{table}
\caption{Comparing MRSD to Human annotated pairs}
\label{tab:mrsd_vs_supervised_pairs}
\centering
\resizebox{0.8\textwidth}{!}{
\begin{tabular}{lcccccr}
\toprule

Method & Aes $\uparrow$ & EA $\uparrow$ & CLAP $\uparrow$ & FAD $\downarrow$ \\

\midrule
Flow-400M                        & 7.58 & 6.90 & 0.34 & 0.78 \\
DPO w. human-annotated 32K pairs & 7.80 & 7.21 & 0.36 & 0.82 \\
\methSmall w. 32K MRSD pairs     & 7.90 & 7.11 & 0.36 & 0.92 \\
\methSmall w. 90K MRSD pairs     & 8.08 & 7.31 & 0.38 & 1.23 \\
\bottomrule
\end{tabular}
}
\end{table}

\subsection{Subjective Preference Per Music Genre}
We conduct an analysis of the subjective performance of \method per musical genre. For the subjective evaluation, we use the open-domain test set, MusicCaps, which cover broad genres and styles. We filter genres with less than 20 answers. We use scores from -2: reference is strongly better, -1: reference is slightly better ... to 2: MR-FlowDPO is strongly better.

Score > 0 implies an improvement w.r.t. the reference model. Results, reported in Table~\ref{tab:genre_analysis}, suggest that MR-FlowDPO improves human preference over all genres that are frequently represented in the test set.

\begin{table}
\caption{Subjective preference per genre: Score > 0 implies an improvement of \method w.r.t. reference model.}
\label{tab:genre_analysis}
\centering
\begin{tabular}{lccr}
\toprule

Genre & Score $\uparrow$ \\

\midrule
all genres & 0.22\pmrtiny{0.13} \\
funk & 0.63\pmrtiny{0.24} \\
other & 0.08\pmrtiny{0.13} \\
rock & 0.14\pmrtiny{0.19} \\
soul & 0.08\pmrtiny{0.44} \\
acoustic & 0.08\pmrtiny{0.32} \\
indie & 0.12\pmrtiny{0.45} \\
hip hop & 0.20\pmrtiny{0.22} \\
lo-fi & 0.43\pmrtiny{0.38} \\
electronic & 0.37\pmrtiny{0.19} \\
jazz & 0.24\pmrtiny{0.22} \\
indian & 0.34\pmrtiny{0.39} \\
pop & 0.27\pmrtiny{0.22} \\
latin & 0.53\pmrtiny{0.34} \\
blues & 0.20\pmrtiny{0.48} \\
ambient & 0.38\pmrtiny{0.30} \\
classical & 0.08\pmrtiny{0.44} \\
metal & 0.27\pmrtiny{0.49} \\
industrial & 0.68\pmrtiny{0.56} \\

\bottomrule
\end{tabular}

\end{table}

\newpage
\section{MusicCaps: List of Textual Prompts used in Subjective Evaluation}

\begin{tiny}
\begin{enumerate}
    \item An acoustic drum is playing along with a bassline giving the song a ska/reggae feeling. The e-guitar strumming on the offbeat supports that feeling while a keyboard is playing a bell-sounding chord. Bongos are setting little accents at the end. A saxophone is playing a melody along. This song may be playing at a festival outside while the sun is shining.
    \item This song contains several drum hits and percussive instruments playing a fast paced rhythm that motivates dancing along. An e-bass is bringing the low end supporting the drums. Cuatro guitars are strumming chords as a rhythmic addition. Trumpets are playing a loud and catchy melody. Some of the musical elements are slightly panned to the left and right side of the speakers. This song may be playing at a cheerful event.
    \item This music is a live instrumental.The music is medium tempo with a dominant piano accompaniment, keyboard harmony, bright drumming, funky bass line, congas rhythm and tambourine beats. The music is a live performance as there are ambient background sounds of people talking and laughing.
    \item This recording contains a dulcimer playing a melody along with loud tablas playing a lot of deep hits that are overloading the recording speakers. At the end a string sound comes underlining with long pad notes. This song may be playing in a live event.
    \item This is the recording of a jazz reggae concert. There is a saxophone lead playing a solo. There is a keyboard and an electric guitar playing the main tune with the backing of a bass guitar. In the rhythmic background, there is an acoustic reggae drum beat. The atmosphere is groovy and chill. This piece could be playing in the background at a beach. It could also be included in the soundtrack of a summer/vacation/tropical themed movie.
    \item This is the live performance of a Turkish folk music piece with arabesque influences. The clarinet is playing the main melody in the lead at a medium-to-high pitch. A secondary melody is playing the qanun in the background. There are supportive elements in the melodic background such as the saz, the keyboard and the bass guitar. The acoustic drums are playing a mellow beat accompanied by rich percussion. There is a vibrant atmosphere in this piece. This could be playing in the background of a Turkish cuisine restaurant or a hookah place. It could also be used in the soundtrack of a Turkish soap opera.
    \item This music is a vigorous electric guitar instrumental. The tempo is fast with an electric guitar rapid melody lead and an electric guitar Accompaniment. The melody is contrasted and syncopated. The music is dark, intense, extreme, dire and compelling Heavy Metal.
    \item This song contains several instruments playing the same melody. Violins; flutes, and tin whistles. An acoustic guitar is strumming chords along and a percussive instrument is playing along with the acoustic guitar. In the background you can hear people talking. This song may be playing in an Irish bar.
    \item The song is an instrumental. The song is medium tempo with a guitar playing solo, steady drum rhythm, groovy bass line and steady bass line. The song is funky and groovy in nature. The song is an ad jingle.
    \item An acoustic guitar is strumming chords. A violin is playing a melody along with a flue and another hammered string instrument. Percussions are being played in a rhythm that is not a 4 bar count. All the instruments are spread across the sides of the speakers. This song may be playing in a desert movie scene.
    \item This song is a sweet duet. The tempo is medium with a melodious, intense piano accompaniment , electric guitar rhythm, steady drumming and synthesiser arrangements. This song is melodic, story telling, spirited, emotional, passionate and sweet. The lyrics are simple and so this song could be a Children’s Song.
    \item These are sounds coming from an advertisement. There are sound effects of bed creaking and laughter implying an intercourse is taking place. Then the act is interrupted by a goofy accordion tune. Sounds from this track can be used in soundboards or in beat-making.
    \item This is a thrilling orchestral piece that feels epic, suspenseful and intense. There are string instruments like violins and cellos, and timpani for percussion. There is one over-bearing vibrational sound that overwhelms the whole arrangement - it is played once and rings for three seconds.
    \item This live recording features an instrumental song. This Regional Mexican song features an accordion playing the main melody. This is accompanied by an acoustic guitar strumming chords. A double bass plays the bass notes. A tambourine acts as the percussion. This song is in an upbeat mood. The song can be played at the entrance of a carnival.
    \item Sounds like happy instrumental karaoke music featuring synthesized horns and a simple chord progression. There is a mallet percussion instrument playing rolls and following a pentatonic melody. Drums playing a simple beat.
    \item The Blues/Pop song features groovy hi hats, punchy snare and kick hits, tinny snare rolls in-between snare hits, addictive brass melody and smooth bass at the very end of the loop. It sounds a bit repetitive, but still addictive and energetic.
    \item This music is instrumental. The tempo is medium fast with a melodious keyboard harmony, steady drumming, groovy bass, synthesiser arrangements , electronically articulated sounds and tambourine beats . The melody is harmonious, pleasant, uncomplicated and well layered. This music is Synth Pop.
    \item The song sounds epic and is fully orchestrated with a string section; brass section; a male choir in the background, fast timpani hits and strings playing a melody. When the song stops, a door knocking sound can be heard along with a short video-game music sequence that does not belong to the previous song. The first song may be playing in a movie trailer.
    \item This song is a mellow Christian hymn in a foreign language. The tempo is slow with soft piano, rhythmic acoustic guitar, steady drumming and keyboard harmony. The music is minimalist and the song is a prayerful, melodious, devotional, emotional and thankful Gospel Song/Hymn.
    \item An acoustic drum is playing a faster groove along with a bassline and a harmonica playing a chord on the offbeat and in a lower key. A e-guitar is playing a melody in the mid range. There are some background noises that remind me of someone brushing his/her teeth. This song may be playing live at a bar.
    \item This music is an Electronic dance instrumental. The tempo is medium fast with punchy drum beats and groovy electronic arrangements.The music is buoyant, energetic, electric, pulsating, youthful and vigorous. The thumpy, rhythmic bass and drumming gives it a groovy dance beat.
    \item This is an opening theme for a TV series. It is an instrumental piece. The main theme is being played by a loud brass section. There is a groovy synth bass line playing. The rhythmic background consists of a strong electronic drum beat. The atmosphere is energetic. This piece could be used in lifting samples for beat-making.
    \item This is an 80s electronic music piece. The rhythmic background consists of a disco electronic drum beat with frequent tom fills. There is a keyboard playing the main tune while a bass and an evolving synth are in the background. The atmosphere of this piece is groovy. This piece could be used at retro-themed nightclubs and parties. It could also be used in the soundtrack of an 80s movie or a TV show.
    \item Very sounds like an old out of tune mechanical music making machine with harpsichord and recorder sounds. Very dissonant and atonal.
    \item This amateur recording features a rock song. The main melody is played on violins. This is accompanied by percussion playing with breaks matching the breaks in the violin parts. At the end, the violins play a sustained note and a banjo starts to play a melody. A tambourine is played on the first count of the second bar after the banjo starts to play. This is an uptempo song and has a happy feel. This instrumental song can be played in a coffee shop.
    \item The song is an instrumental. The song is medium tempo with a flute paging embody, various stringed instruments playing in harmony and no percussion. The song is passionate and emotional. The audio quality is average .
    \item Someone is playing a track from speakers. This song contains a strong e-bass playing a funky bassline along with a funky drum groove. Then a piano comes in playing a jazzy melody in one scale accompanied by a synth brass sound swelling into existence and playing a short rise before leaving again. This is an amateur recording but of decent audio-quality. This song may be playing in a jazzbar.
    \item This is a live acoustic performance piece. The didgeridoo is being used in varying ranges and effects to play many different tunes and to get a techno-like sound variety. There is a whole bunch of percussive instruments being used from membranous percussion to bell percussion. It all adds up into one big sensational performance. There is an eccentric atmosphere. This piece could be used in the soundtrack of a movie with a primitive/jungle setting. It could also be used in action or pursuit scenes in movies.
    \item This music is an energetic Rock and Roll instrumental. The tempo is fast with animated, infectious drumming, lively trumpets and tambourine beats. The music is upbeat, pulsating, thumping, vivacious, bright, with a perky dance groove. This music is Rock and Roll.
    \item This is an instrumental heavy metal piece. There is a bass guitar gently playing the melody with a clean sound. It is a melodic piece with a touch of melancholy. This piece can be used in the soundtrack of a drama movie or a TV series, especially during flashback scenes. It could also be used in video games during story exposition cutscenes.
    \item This song is an instrumental. The tempo is medium with a melodious keyboard harmony, rhythmic guitar accompaniment, punchy drumming, subtle bass line, synth arrangement and tambourine beats. The music is soft, mellow, ambient, pleasant, uplifting, and mellifluous.
    \item The song is an instrumental piece with various tracks. The song is medium tempo with a piano accompaniment, groovy bass line , steady drumming track, a male sighing in pleasure and water filling a bottle tone. The song is a mix bag of music tracks for various scenes. The song is an instrumental track for a funny YouTube video.
    \item This is the recording of a tuning video in the key of D sharp. There is an electric guitar playing the note on each string repeatedly. This piece could be used to lift samples to be used in a beat or in the making of a virtual guitar plug-in.
    \item The clip is an upbeat, groovy funk song with jazzy elements. The retro vintage electric piano chords add an old school jazzy feel to the song. The bassline is intricate and bouncy and enables easy dancing.
    \item The Electro Pop song features a buzzy sustained synth lead, "4 on the floor" kick, short reversed snare hits, claps, shimmering hi hats, groovy piano chords, sustained synth pad chord and arpeggiated synth melody. It sounds energetic and like something you would hear in clubs.
    \item A accordion-like wind instrument is playing a melody along with someone playing a rhythm on the tablas. This song may be playing at a music school jamming with the teacher.
    \item This music is a lively fiddle instrumental. The tempo is fast with a percussion accompaniment. The music is spirited, vibrant, enthusiastic,cheerful, happy , merry and sunny. This music is Country Folk.
    \item This song is an instrumental. The tempo is fast with a lively keyboard harmony and rhythmic steel pan . It is minimal and rhythmic. The harmony is simple , straight and the steel pan adds an interesting layer to it. The music is engaging, interesting, energetic, spirited and melodic.
    \item The song is an instrumental music piece. The song is slow tempo with an electric guitar playing lead with heavy echoes, steady drumming and a keyboard harmony. The song is casual and slightly out of rhythm. The song is an instrumental  with a casual listening vibe.
    \item This music clip is an percussionary instrument. The tempo is medium fast with steady bass drum, energetic snare drum beat and cymbal rides. This music is a youthful, punchy, energetic, simple drumming style.
    \item This song is an instrumental. The tempo is fast with an animated keyboard harmony, funky bass lines; groovy electric guitar melody, slick drumming and synthesiser arrangements. The instrumental is upbeat, catchy, vibrant, animated, youthful and vivacious. This song is an Electro Rock/ Synth Rock.
    \item This is a mellow, groovy hip hop beat. Its distinguishing factor is a vinyl scratching technique used by DJs. It also consists of what sounds like a brass instrument and forms a chord progression along with the bass that plays at the same cadence.
    \item The song is an instrumental. The tempo is slow with a piano accompaniment and a strong bass line, with people clapping , whistling, and cheering. The song is being played at a live arena with people getting excited to the song introduction.
    \item This is a live DJ performance. There is a groovy bass line and a mellow electronic drum beat in the rhythmic background accompanied by the tune of a melodic guitar sample. The turntable is used to make a scratching sound over this track. There is an urban feeling to this piece. It could be used in the soundtrack of a crime movie/TV show taking place in the big city. It could be used in the background of an interesting sports video.
    \item The song is an instrumental. The tempo is slow with percussive instruments playing, bells percussions and a music box melody. The song is educational and designed for toddlers. The audio quality is pretty average.
    \item A female and male duo yodeling over a fast bluegrass band. The bass and drums are steady and simple as is the chord progression. The guitarist and other instrumentalists are virtuosic.
    \item This is a pop-sounding jingle at the background of an instructive video. There is an electric guitar playing the main theme while the bass line is being played by a synth bass. In the rhythmic background, a very basic 4/4 beat is being played on the electronic drums. A ping sound effect signifies the end of the instruction in the video. The atmosphere is very generic and common-sounding. This piece could be used as a jingle in advertisements and tutorial videos.
    \item This is an orchestra playing Arabic music. The melody is played by the conjunction of multiple instruments such as violin, flute and qanun while tambourines create the percussive background of the piece. It has an oriental atmosphere that could be very fitting in a movie or a show taking place in the Middle East. It could also be a good background music for Arab cuisine restaurants.
    \item This song is a choral harmony. The tempo is slow with intense bass and an instrument that sounds like a viola. The song is grim, grave, intense, sinister; eerie; spooky; scary, chant-like, and dreary.
    \item This audio clip is an instrumental. The tempo is low with a subtle violin harmony, keyboard arpeggio and synthesised organ. The music is lilting, grim,incessant, intense, sad, bleak, enigmatic and ambiguous. The music sounds like the background score for a movie.
    \item This song contains a string section playing a short stroke melody in the lower range along to a digital acoustic drum playing a rather complex groove. Acoustic guitars are playing a melody. This song may be playing in a heroic movie scene.
    \item This song is an instrumental. The tempo of the background music is slow , with electric guitar harmony and rhythmic drum with loud chaotic, dissonant sounds. The music is out of sync as the unidentifiable sounds are accelerated.
    \item This instrumental song features a didgeridoo being played by the vibrating of the lips. This features a continuous droning sound inter-spread with bass sounds. This is accompanied by a handpan playing the percussion and backing notes. This song can be played in a movie featuring a tribal initiation ritual.
    \item This is a marching band piece performed by an orchestra. There are trumpets playing the main theme while tuba and cello are playing the bass notes. There is also a constant string part holding the root notes of the melody. The snare drums are playing an accentuated, militaristic drum line. There is an epic atmosphere that gives a call-to-action feeling. This piece could be used in the soundtracks of action movies and war movies in particular. It could also be used in the soundtracks of video games from the action and first-person-shootout genres.
    \item Someone is playing an e-guitar with different effects over an acoustic drumgroove. Some of the sounds are panned to the right and left side of the speakers. This song may be playing at home practicing guitar.
    \item The song is an instrumental. The tempo is medium with a guitar accompaniment and leads into a strong drumming rhythm and bass line. The song is exciting and energetic. The song is a movie soundtrack for a travel show or a documentary.
    \item Trio recorded live in a concert hall with a loud upright bass, piano and acoustic guitar playing a solo. Jazz feel and chord progression with no drums. Heartfelt performance.
    \item This song comes from a music box. The backing melody is repetitive which is played in a set of three notes followed by a set of two notes. The main melody is high pitched and sounds like bells. There are no other instruments in this song.
    \item This is a meditation piece. There are waves in the background sounding similar to sine waves. There is an evolving synth signifying a passage. There are various sound effects, especially in high pitch. There is a hypnotizing aura. This piece could be used in the background of a meditation video.
    \item This song is a romantic duet.the tempo is medium with a soft keyboard harmony, steady drumming, violin harmony, chimes, Cajun beat, bongos rhythm, steady bass line and acoustic guitar .the song is a proposal song.it is romantic, emotional, passionate, soft, ambient, mellifluous and dulcet. This song is a Soft Pop song.
    \item The song is an instrumental. The song is medium fast tempo with a bass guitar soloist playing a cool accompaniment groove with no other instrumentation. The song is passionate and energetic. The song is a bass guitar demo or a home music video.
    \item The instrumental music features a piano playing a romantic song. A group of strings accompany the pianist with warm harmonies. The overall atmosphere is romantic and touching.
    \item The Classical song features wide sustained strings and subtle mellow piano chords. It sounds sad and emotional - like something you would hear in the video games, as a background music for a sad scene.
    \item This is a live recording to which music was later added. The main audio in the clip includes a squeaky toy being repeatedly pressed. The song that was later added is a slow ballad that includes piano and violin.
    \item This music is a flamenco piece. There are two acoustic guitars, one playing the leading arpeggio melody and the other creating the rhythmic background by strumming the chords. The characteristics of the song makes it clear that it is influenced by Spanish music. It could be used in a thriller movie soundtrack or as a flamenco dance accompaniment piece for dancing courses.
    \item The song is an instrumental tune. The song is medium tempo with a steady drumming rhythm, groovy bass line, trumpets playing, brass band section and keyboard harmony. The song is exciting and energetic. The song is a tv show tune and has fun jazz roots.
    \item Sounds like the intro to a children's program. Children's music featuring vibraphone, a children's choir and old tape static in the background. It has a happy, bright feel.
    \item A live recording of an instrumental piano trio. The recording is boomy and low fidelity. The drumming is active and the piano playing is pretty. The drum kit features a sidestick. Electric bass is also present. Crowd noise can be heard. Sounds like gospel music that might be heard at a church service.
    \item The music features a melody being played in unison by a piano and a group of strings. At the same time these two also fill up the music space with harmony. A type of drum that resembles the taiko drum but I think is smaller in size can be heard in the background. In the background a synth pad provides soft harmony.
    \item The song is upbeat, groovy and jazzy. The main percussive elements are created by live tap dancing on a stage, as well as more base oriented percussion from a cajon drum. There's a grand piano playing an intricate melody.
    \item This music is instrumental. The tempo is fast but the music is muffled as it is superimposed by a loud buzzing sound along with some tapping,scuffling, finger snapping ans tapping sounds. The music sounds like energetic Electronic music with vigorous drumming.
    \item This music is an electric guitar solo. The tempo is medium fast with an animated electric guitar riff without any accompaniment. The music is loud, powerful, rhythmic and youthful with a bit of static or white noise .
    \item This is a Nepali music piece. There is a string instrument and a local flute playing the melody. The rhythmic background is provided by percussion. It has a vibrant, dreamy feeling to it. This piece could be used in a dream sequence of a movie or TV show taking place in South Asian countries.
    \item This is a chiptune piece. The only instrument being used in this track is a granular synth playing the melody at a medium-to-high pitch. It is an analog sounding piece. There is a nostalgic atmosphere to it. This piece could be used in the soundtracks of cartoons and arcade video games.
    \item This folk song features a reed instrument playing the main melody. A similar melody is played on a percussion instrument by a skilled percussionist. Another percussion instrument plays a tabla-sounding instrument. Cymbals also play a melody keeping time with the melody being played. The end of the song features an ascending run on the reed and the harmonized percussion.
    \item This song contains a digital drums playing a four on the floor groove with a big kick sound on every beat. A e-guitar is strumming rhythmic funky chords. Towards the end you can hear a synth string playing in a high register. This song may be playing for a advertisement.
    \item This song is a percussion instrumental. The tempo is fast with intense and rapid drum rhythm along with sound rhythmic clashing of cymbals with sound of clapping in the background and a man grunting. The music is animated,vigorous ,energetic, and enthusiastic.
    \item This song contains mallet instruments playing a fast melody in the mid and high register along with low notes as bassline. The marimba is full of reverb. Then a bass drum comes in playing on every beat. This song may be playing in an advertisement.
    \item This is a Bali folk music piece. The instruments that are used are mostly percussive in nature traditional to the music of Bali such as gamelan, kendang and gong. There is a hypnotic atmosphere in this piece. It could be used in dream/hypnosis sequences in the soundtracks of movies and TV shows.
    \item This music is an enthralling Sitar instrumental. The tempo is fast with an animated and melodic Sitar lead Harmony and tanpura accompaniment. The music is based on Hindustani Classical notes. It is captivating, intense, mellifluous, engaging,and fervent.
    \item This song is a spirited instrumental. The tempo is fast with enthusiastic brass harmony , with trumpet flourish, trombone and violins, cello and viola playing an upbeat melody along with vigorous drumming, piano accompaniment and cheerful, animated choral backup. The song is animated, vibrant, energetic and full of life. The sound of the hooves of horses and whistling indicates that this is Country music.
    \item This music is an electronic instrumental; the tempo is medium with a solo electric guitar lead and no accompanying instruments. The music is melodic, rhythmic, and groovy.
    \item An acoustic drum set is playing a shuffle groove with hits on the ride along with an e-bass playing a blues walking bassline. Someone is playing a slide-e-guitar. This song may be playing at a live concert.
    \item A steeldrum-orchestra in playing a composition containing bass sounds, mid range chords and melodies in the higher register. This sounds soothing. This song may be playing as a live performance.
    \item The music is purely instrumental and it features two acoustic guitars. One of them is playing a melody in the high register while the other one is providing accompaniment by strumming chords.
    \item Big band jazz style music featuring a lead clarinetist soloing over a walking bassline, electric jazz guitar, a wind section, brass, string section, and a choir.
    \item The music features a piano playing short interventions/melodies. A bass guitar (could be synth too) plays long notes. In the background, but very noticeable, is an electronic sound playing one long note throughout the music excerpt. The overall atmosphere is introspective.
    \item Someone is playing an acoustic guitar finger picking a melody. On top of that, a distorted e-guitar is playing a solo melody with delay on it. This song may be playing at home creating a song in a DAW.
    \item This is an Arabian themed instrumental music piece. The tune is played by a virtual eastern-sounding instrument while there is a virtual strings backing for the bass notes. The rhythmic background consists of a repeated percussion beat. There is an intriguing feel coming from the back-and-forth movement of the tune. The piece could be used in an eastern historical drama.
    \item This is an electronic dance music piece. The tune is being played with a melodic bell preset. There is a medium-pitched synth holding the chords in the background. The rhythm is provided by a crisp electronic drum beat. The atmosphere is vibrant. This piece could be playing at a nightclub. It could also be included in the soundtrack of a movie taking place in winter.
    \item This is a dance music compilation. There is a variety of music pieces ranging from pop to electronic to folk. The sounds made by the shoes of the dancers can also be heard. The atmosphere is rather random. Parts of this recording could be lifted as samples to be used in beat-making.
    \item The song is an instrumental. The song is medium tempo with steel pan playing melody, guitar strumming and steady percussion rhythm. The song is exciting and calming. The song is a live band performing tropical Caribbean tunes.
    \item This music clip is an instrumental with gamelan instrumental lead along with a hand percussion. The tempo  is medium with melodic sounds and harmony from the striking of metal bowls or plates. The music is soft, melodious, calming, engaging and interesting and sounds a lot like a xylophone.
    \item This song is a pleasant love song duet. The tempo is medium with acoustic guitar and ukulele accompaniment.the song is mellow, soft, emotional, romantic, nostalgic. This song is Country/Folk pop.
    \item The instrumental latin music features a big band. The trumpets, saxophones and trombones play the same melody but on different pitches. The percussion section includes a drum kit played in the latin style and a pair of congas. A piano accompanies the brass section and blends in with the rest of the band. An electronic sound can be heard playing the same note throughout the music excerpt.
    \item This is a classical music performance. There is a gentle piano tune playing in the background while there is a theremin playing the main melody. The performance has a unique feel to it. The atmosphere is out of the ordinary but also heart-touching. This piece could be used in the soundtrack of an animation movie/TV series.
    \item A blues rock power trio with electric bass and overdriven electric guitar playing a riff together and the drummer playing a backdoor shuffle. It is instrumental with a swung rhythm.
    \item This music is instrumental. The tempo is slow with a flute playing a melodic lead harmony and later joined by a rhythmic acoustic guitar. The song is dulcet, warm, peaceful; calming, meditative,pensive, contemplative, deliberate, mellifluous and euphonious. This music is an Indian Classical.
    \item The music features an electric guitar playing what sounds to be a solo over a backing track or an actual recording of a song. The notes played in this passage are played fast. The vocabulary used by the guitarist resembles that of the jazz-fusion genre.
    \item An e-guitar takes the fast paced lead melody accompanied by another guitar strumming chords on the offbeat. The drums are playing a rhythm that motivates me to dance. This song may be playing somewhere on a street-performance in an Asian country.
    \item The track features a cinematic ambience with different sound effects like waves or earthquake sounds. In the foreground there's a bright piano playing a melody with a positive vibe. The track sounds like an extract from a video game,
    \item This song is a lively Jazz instrumental. The tempo is medium with a spirited piano harmony, syncopated drum rhythm and groovy bass lines. The music is ambient, lively, engaging and catchy with complex chords, irregular beats, improvisations , syncopation and creative freedom for improvisations.
    \item This music is a percussion solo. The tempo is fast with the drummer  playing an energetic solo. The beat is vigorous, enthusiastic, spirited, infectious and punchy with a flourish of cymbals.
    \item The song contains a flutesound, brass-section, e-bass and some lower register keys all playing the same funky and repetitive melody. The drums are holding a straight rhythm while the hits on the ride are played in a light swing. This song may be played at a rollerblade disco.
    \item This music clip is a loud, boomy drum beat. The tempo is slow with the powerful and rhythmic beat of the Big Chinese drum. It is loud, resonating, vibrating and powerful and sounds ceremonial, celebratory and royal and used for announcements.
    \item This is an instrumental live downtempo music piece performance with deep house elements. There is a synth bass lead while the medium-pitch strings hold the melodic background. An occasional piano riff can be heard every now and then. There is a mid tempo electronic drum beat in the rhythmic background. It has a very groovy mood. This piece could be played during earlier hours at a nightclub. It could also be used in a tech device advertisement.
    \item This music is a violin instrumental. The tempo is medium fast with the violins going from lively to soft with a romantic piano melody and acoustic guitar accompaniment. The music is orchestral with a dulcet violin symphony. It is pleasant, mellifluous, passionate, romantic; euphonious, emotional, sweet and engaging. This music is a contemporary orchestra .
    \item This is a clip containing vinyl or deck scratching. The DJ manipulates the deck back and forth to create scratching sound effects which have a melodic and percussive type of vibe. There's a mellow synth on which a simple progression is being played, and a simple hip hop drum beat is being played.
    \item This is an instrumental piece carrying the characteristics of Carnatic music. An acoustic guitar is being played sideways to reach the musical effect of a tambura. The piece could be used as a soundtrack in movies/shows that take place in India.
    \item This is an upbeat swing style song with a choir of brass instruments playing a vibrant motif. The motif is a sort of crescendo that announces the next segment. It would be suitable as an intro to the arrival of a character in a musical or movie.
    \item A full orchestra is playing a slow dramatic, emotional and soft composition. In the background whispering can be heard. This is an amateur recording. This song may be playing at a theater dancing ballet.
    \item This slow blues song features a solo played on a guitar. The tempo of this song is slow. This is accompanied by percussion playing a simple beat. The bass follows the 12 bar blues pattern. The piano plays chords in the background. The mood of this song is romantic and the pattern played on the guitar is similar to a call-response theme. This song can be played in a bar.
    \item A song is played in the background containing a brass melody. In the foreground you can hear a lot of clicking noises that do not belong to the music. This is an amateur recording. This song may be playing in a casino.
    \item This music is a Country song instrumental. The tempo is medium with a pedal steel  guitar lead harmony with steady drumming and keyboard accompaniment. The music is a simple,minimalist, sweet, dulcet, sentimental, nostalgic, soothing, pleasant and mellifluous , classic Country music instrumental.
    \item The music is purely instrumental and it features a 12-string acoustic guitar playing the same note over and over. Since it's a 12-string guitar the note can be heard played twice really fast. I guess a better name would be tuning instead of melody, because it sounds as if the guitarist is trying to tune the 2 strings to play the same note together.
    \item Audio of a percussion instrument playing by itself. Sounds like a tambourine, riq, or pandeiro, playing a steady beat, alternating between a cymbal sound and a bongo drum sound.
    \item This song is an instrumental. The tempo is low with a xylophone like instrument harmony, sound of clock ticking and papers rustling . The music is spooky, eerie, suspenseful and sinister. This song is Pop.
    \item The song is an instrumental. The song is medium tempo with a steady drumming rhythm, cymbals crashing, piano accompaniment and a xylophone playing a cool melody. The song is emotional and passionate. The song is an ad jingle for a technology solutions company.
    \item The song is instrumental. The tempo is medium with sitars and other stringed instruments playing in unison, tabla percussions and a harmonium playing the lead. The song is a classic hindustaani instrumental.
    \item The Rock Orchestra song features energetic crash cymbals, punchy kick and snare hits, groovy bass, wide electric guitar chords and simple strings melody. It sounds epic, powerful and energetic - like a theme song for an opening track of some anime.
    \item This is a live recording of an instrumental electronic music beat. The rhythmic foundation of the beat consists of a simple medium tempo electronic drum beat. There are sounds coming from the background resembling a restaurant. It is a minimal-sounding groovy beat that is also danceable. With the removal of ambient sounds and a clean mix, this track could be used in the soundtrack of a TV show with an urban setting.
    \item This is an instructive flamenco guitar piece. The player is depicting how the piece must be played by starting it slow and eventually getting faster. The theme is repeated by the player.
    \item This is a loud heavy metal piece that is played as a soundtrack for a video game with sound effects of fire and a dinosaur roaring in the beginning. There is a distorted electric guitar playing a simple tune while loud acoustic drums provide a heavy metal beat as the rhythmic background. The theme has a very violent feel to it.
    \item The Corporate instrumental contains wide plucked strings, simple bells, punchy, roomy kick hits and soft rimshots. At the very end of the video there is a shimmering tambourine. Judging by the silence at the end, it could be said that this is an outro of the jingle. It sounds happy, fun and uplifting, like any corporate sound should sound like.
    \item This is a clip of a piece on the timpani orchestral percussion instruments. The drummer repeats the same motif twice. The sound of the drums is regal and grand.
    \item This music is a pleasant instrumental. The music is slow tempo with a soft cello and guitar duet with a whistle melody. The song is mellow, soothing, calming , dulcet, warm and relaxing in contrast to the background noises of people talking and walking. The musicians are buskers.
    \item This is an intricate country guitar performance. The guitarist plays slide guitar. The playing is complex. There is a tambourine as a percussive element on every beat. It's a live recording.
    \item The song is an instrumental. The song is medium tempo with an overdriven guitar playing a jarring lead in rhythmic fashion and guitar over tones that sound like bell tones. The song is aggressive and experimental. The song is a modern rock intro.
    \item This is an instrumental lounge music piece. There is a vague melody with reverberations of an ambient synth and occasional chord strumming from an electric guitar. The electronic drum hits are spread out at a low tempo. The general feeling of this song is chill yet also trippy at the same time. This piece could be used in a technological device release video. It could also be used in scenery shots in a movie or a TV show.
    \item This song contains a lot of digital orchestra instruments. Celli is playing a short bowed stroke chord providing rhythm along with  other cello sounds being picked in the mid range. A flute is playing a melody on top. Other string instruments are playing a plucked arpeggio melody. While timpani and cymbals are adding hits. The instruments are spread across both sides of the speakers. This song may be playing in an arcade/adventure video game.
    \item This is a gear showcase jam. The only instrument that is being played is a clean sounding electric guitar playing a mellow solo. A delay effect pedal is applied on the guitar. The atmosphere is calm and dreamy. This piece could be used in an advertisement jingle. It could also be playing in the background at a rock bar.
    \item This is an amateur recording of a guitar solo for a rock music piece. The solo is being played on an overdriven electric guitar while the piano and the bass guitar in the backing track play the main tune. There is a 6/8 rock beat played on the acoustic drums for the rhythm section. There is a raw feel to it. The piece can be used to gather electric guitar samples to be used in beat-making.
    \item This is an instrumental progressive rock music piece. There is an electric guitar playing complex tunes and chords with a pitch shifting effect. There is a psychedelic feel to this track. Parts of this recording could be used in an advertisement jingle.
    \item Percussions are playing together with a clarinet that takes the lead melody with very long notes. A bowed instrument is playing along with little fill-ins. An electric bass is playing a rather funky groove. The whole song sounds like it was made for dancing joyfully.
    \item This music is an intense instrumental. The tempo is medium with hard hitting drums, ceremonial big drums, amplified piano and violin harmony. The music is syncopated and intense. It is like a background score for a movie.
    \item Solo ragtime piano playing a lilting instrumental section with a bouncy bass part in the left hand. The bell sound effect happens once. Good for a sunny day at a County fair or carnival.
    \item Someone is playing a melody on an acoustic guitar and someone another melody on a oud. The two melodies complement each other. This song may be playing in a live presentation.
    \item This song is an instrumental. The tempo is slow with vigorous drumming, keyboard and electric guitar harmony,beatboxing and the sound of a car horn .The song is vibrant, punchy, vigorous and upbeat with a dance groove. This song is a Pop Hit.
    \item This song contains a string instrument playing a fast and complex melody/solo while percussion instruments with a deep kick drum sound, shakers and complex clapping create a solid foundation. This song may be playing in a movie-scene in the desert.
    \item This kids song features groovy synth keys, chords, simple piano melody, echoing and wide pig oink sound effects, shimmering shakers, muffled claps, soft kick hits and groovy bass. It sounds fun and happy - like something that children would listen to in their favorite tv shows.
    \item This clip is an instrumental. The tempo is slow with the sound of an acoustic guitar being tuned. This is a tutorial on how to tune a guitar.
    \item The Electro song features a resonant synth melody, repetitive synth melody and widely spread, mellow synth pad chords playing. It sounds relaxing and calming, like something you would listen to while chilling.
    \item This is an excerpt from a cartoon that involves sound effects resembling an intro to a minimal techno song. There is a breathing sound coming from one of the characters. There is also a glass sound effect. There is an overarching sci-fi character to the sounds.
    \item This music is an electric guitar instrumental. The tempo is slow with an electric guitar lead and harmony of a popular song. No other instrument is used. It is mellow, pleasant, nostalgic, soft , mellifluous and soothing.
    \item The song is an instrumental. The song is medium tempo with a groovy bass line and a guitar synth tone playing melody. The song is exciting and adventurous in nature. The song is a video game song track.
    \item This ambient song features synth pads playing in the background. Synth sounds are layered to give a wave-like feel. There is no percussion in this song. Bells are played at intervals. Toward the end, a mid-range flute sound is played. The sound of rattles are played in the background. This song has an ethereal feel. This song is meditative and can be used in a meditation or yoga session.
    \item This instrumental is a Heavy Metal instrumental. The tempo is fast with hard hitting drumming, furious and vigorous amplified keyboard playing a harmony, electric bass guitar and electric guitar accompaniment. The music is intense, grim, compelling, passionate, powerful and harmonious. The vibe of the music is serious, sinister, grim and steely. This is used in Hard Rock/Heavy Metal.
    \item This is an amateur low-resolution recording of an electric guitar with the chorus pedal effect. The player keeps on strumming chords to show the effect of the pedal. This piece could be used as an old-fashioned rock guitar sample for a beat.
    \item Someone is scratching along with a backing track that contains a strong kick, a snare and a dark sounding melody. This is an amateur recording. This song may be playing at home practicing scratching.
    \item This music is instrumental. The tempo is slow with an electronic keyboard playing a simple, high pitched harmony. There are sounds in the background like whirring, whooshing and clicking. The music is not very clear as the audio clip is muffled, but it sounds like video game music or the music Ina child’s electronic toy.
    \item The track is instrumental. The tempo changes building suspense, violins trilling, string section harmony, brass section plays percussively and piano plays suspenseful notes. This is a movie soundtrack. The audio quality is average.
    \item This song contains someone playing a bass-ukulele along to a piano playing jazzy chords along with a soft jazz piano. The main melody is being played by an alto saxophone. This song may be playing in a jazz lounge bar.
    \item This song is an instrumental. The tempo is medium with lively trumpets, trombone ,enthusiastic percussion like the snare, cymbals and bass drum, tambourine , an intense and bright string harmony of violins, viola and cello. The music is intense, serious and is gradually building up. This is an Orchestral piece.
    \item Someone is playing an acoustic guitar along with someone playing a solo melody on top with another plucked string instrument. They are accompanied by someone playing a djembe or another percussive instrument. This song may be playing sitting around a bonfire.
    \item This is a classical music piece performance. It is meant to be played at a wedding. There is an organ playing a simple and festive tune. There is an atmosphere of celebration in this piece. This music would fit perfectly with weddings. Organ samples could also be lifted from this recording to be used in beat-making.
    \item This is a recording of two didgeridoos. They are playing low notes and create a very low vibrational tone. There is a melody, but it is not easily recognizable.
    \item Live performance of a girl band group, featuring surf guitar and a go-go beat that transitions to modern pop production with loud bass and electronic drum breaks. Audio is low fidelity. Would be popular with teenage girls.
    \item This song contains a evolving synth pad sound in the higher register with noise. A piano is playing a sad minor chord progression in the mid range. Then a strong industrial sounding digital drum comes. This song may be playing in a very sad movie scene.
    \item A drum is playing a lot of crash hits with fast and accurate snare rolls along to a backing track that contains an e-guitar playing a riff in a lower key. This song may be playing at a studio practicing.
    \item An acoustic piano is playing a sad sounding composition while someone is playing a synthesizer lead-sound that sounds like someone crying. You can hear a lot of white noise in the recording. This seems to be an amateur recording. This song may be playing on a TV show.
    \item A sample of an alto saxophone playing a repeating melody is being played. At the end some percussive hit comes in with a riser. The whole song sounds upbeat and invites the body to pulse. This song may be playing at a festival with a DJ on stage.
    \item This is a live recording from a rock music concert. There is an electric guitar playing the rhythm section upfront with a loud drum background. A harmonica solo is being played over the guitar and the drums. The song carries a playful and energetic atmosphere. This music could be played at a bar, especially a rock bar.
    \item This is an electronic DJ set in the style of Jamaican reggae music. There is a DJ performing over a track that has a groovy beat being played by the electric guitar and the electronic drums. There is a tropical atmosphere. Samples could be lifted from this recording to be used in the beat-making.
    \item The song is an instrumental piece. The song is medium tempo with a furiously playing timpani percussion, string section playing, horns playing and low pitched harmony that evokes danger and suspense. The song is a modern western classical instrumental.
    \item This is an instrumental heavy metal piece. The distorted electric guitar is playing an aggressive riff while accompanied by a bass guitar. In the rhythmic background, there is a slow but hard-hitting heavy metal beat played on the acoustic drums. The atmosphere is loud and violent. This piece can be used in the soundtrack of a movie/TV show involving crime. It could also be played at a rock bar or a sports venue.
    \item This is a new age piece. There is a flute playing the main melody with a lot of staccato notes. The rhythmic background consists of a medium tempo electronic drum beat with percussive elements all over the spectrum. There is a playful atmosphere to the piece. This piece can be used in the soundtrack of a children's TV show or an advertisement jingle.
    \item This is a contemporary classical music performance. The piece is being played on the grand piano with an accentuated playing style. There is a lot of emphasis on the notes. The atmosphere is dramatic. Parts of this piece could be included in the soundtrack of a documentary. It could also be used in the soundtrack of a mystery/horror video game.
    \item Someone is playing a loud melody on a steeldrum along to a backing track with a lot of percussion and an upright bass. This song may be playing at a beach concert.
    \item This song is an instrumental. The tempo is medium with a melodious lead on the electric guitar, steady drumming, groovy bass lines and acoustic guitar accompaniment. The melody is simple, mellow, pleasant, melancholic, nostalgic and ambient. This instrumental song is a Classic Rock and Roll.
    \item Calming ambient synth pads and synthesized bell sounds combined with ocean sounds and bird sounds. Meant to inspire tranquility. No drums.
    \item These are sounds from a movie. There is a metal object falling to pieces on the ground. There is a cinematic string sequence that gives a dramatic atmosphere to the track. There is also a feeling of suspense. This could be playing in the soundtrack of an action-filled movie.
    \item This is a live performance of a classical music piece. There is a grand piano playing dramatically in minor keys. The piece has a grandiose and mysterious aura to it. This piece could be used in the soundtrack of a horror movie. It could also be used in a horror video game theme.
    \item This music is instrumental. The tempo is medium with atmospheric synthesiser, soft keyboard harmony and subtle bass lines in the background. The music is progressive with a lot of digital sounds giving it a psychedelic,meditative, trippy, trance like vibe. This instrumental is Synth Rock/Progressive Rock.
    \item A synthesizer is playing a loud and wide pad sound that gets modulated while a soft and deep kick is playing every beat. This is an amateur recording. This song may be playing giving someone a tutorial on a synthesizer.
    \item The children's music sounds as if it's being played from a child's toy and recorded with a phone. The music starts playing after a rattling noise is being made and stops after a while to be activated again by the rattling sound.
    \item This music is an Indian Classical Instrumental. The tempo is medium fast with an ensemble of a lively violin, melodic flute and rhythmic accompaniment of the Indian percussion gadam and mridangam . The music uses ragas, talas and Sruti to form a harmony. It is classic, lively, rhythmic, engaging and enthralling.
    \item The music features a melody-less section, just the accompaniment. A piano is playing chords at the beginning of every measure. The kick drum plays on beats one and three and the snare drum, which seems to have some beads on its membrane, on beats two and four. An electronic sound is playing the same harmonies that the piano is playing. Two distorted sounds can be heard in the background, one of them most probably coming from an electric guitar and the other from a synth bass.
    \item The song is an instrumental. The song is medium tempo with a groovy bass line, steady drumming rhythm, guitar accompaniment, saxophone playing melody and bells percussions. The song is cheerful and festive. The song is a Christmas song with a lot of fun and cheer.
    \item The Regional Mexican song features solo flute melody over wooden percussive elements, groovy piano melody and groovy bass. It sounds fun, happy and it is uplifting and energetic - like something you would dance to in some latin bar.
    \item This is a techno music piece played in the background of a car video. The sounds coming from the car such as the exhaust sound can be heard in the video. The electronic music piece consists of a crisp synth sound playing a repeated theme while there is an electronic drum beat in the rhythmic background. There is an aggressive atmosphere in the piece. Samples lifted from the track could be used in the soundtrack of a car racing video game.
    \item The music features a stringed instrument playing a melody. I'm not sure what instrument it is, it sounds oriental. I can say for sure it's not an acoustic guitar. This instrument has a synth delay effect on, meaning whenever a note is played, a device plays that same note with the sound of a synth.
    \item A nostalgic feeling trip hop song with an off kilter, Dilla inspired drum beat, a jazz piano playing complex chords and a male rapper.
    \item A cheery ukulele ensemble featuring ukulele strumming and a harmonized ukulele melody.
    \item A guzheng is playing a traditional melody with some notes bending while a flute is playing notes in a tremolo. In the background a string instrument is played with a bow. All the instruments are full of reverb and delay. Also are they panned to different directions of the speakers.
    \item This song may be playing at a ceremonial dance or theater.
    \item The song is an instrumental improvisation. The song is medium fast tempo with a groovy bass line, electric guitar playing, saxophone playing, jazz drumming with percussive hits and a xylophone playing. The song is highly improvisational and energetic. The song is a modern jazz fusion or improvisation jam.
    \item Violins are playing a lead melody underlined by celli playing supportive chords. A brass section is playing a countermelody while shakers and other instruments are providing rhythm. This song may be playing at a folkfest with a dancefloor.
    \item This piece is an instrumental. The tempo is fast with hard hitting percussion, cymbals clanging, clapping, rhythmic chanting, and shakers. It is vigorous, spirited, vibrant, energetic,intense and engaging.
    \item Someone is playing a snare with brushes along to zitars playing a repeating melody. In the background you can hear an e-bass and an acoustic guitar strumming chords. This song may be playing in a live concert.
    \item This music is an acoustic instrumental. The tempo is medium with an acoustic guitar lead harmony, with minimalist instrumentation. The music is soft, soothing, pleasant, engaging, and delicate.
    \item This audio contains sound effects like horn honking and mallet instruments walking up the scale. This may be playing on a machine in a casino.
    \item The performer is snapping his fingers in rhythm with the upbeat japanese music playing in the background. The song is a j-pop song and features vibrant rhythmic synth activity and has a general dance feel to it. It's a live recording.
    \item This slow pop song features a synth playing the bass notes as well as the lead melody. The music is in 6/8 time signature. In the first bar, two chords are played. The main melody is played in the second bar. The synth accents the notes on a higher register. The bass plays at off timing. The first two notes of the bass are played on the first count and fourth count of the first bar. The next bass note is played on the first count of the second bar and the following note is played on the 'three and' count of the second bar. The mood of this song is relaxing. This song can be played in a coffee shop.
    \item This music is a lively xylophone instrumental. The tempo is medium with minimal instrumentation of a xylophone and piano harmony. The music is pleasant, melodic, quirky,a little peculiar and unusual like music in a circus for a clown act. There is also a sound of blowing or shushing in the background.
    \item This instrumental song features a deep bass sound that is playing continuously in the background. A piano sequence is played which gives this song an eerie and suspenseful feel. A layer of synth swells are played when the piano sequence is played. At the end, high pitch sounds are played in the background at a low volume as if the sound source is far away from the mic. The mood of this song is suspenseful as if there is impending danger. This song can be played in the lounge of a video game.
    \item A high pitched long note is playing along with other synthesizer pad sounds to create a bizarre and  mildly creepy atmosphere. A synthesizer lead is playing a disharmonious melody. The sounds are full of reverb and delay. This song may be playing in an alien arrival movie scene.
    \item This is a gear showcase jam. The only instrument being played is an electric guitar being strummed lightly. There is a chorus ensemble effect on the guitar making it sound like multiple guitars are being played at the same time. There is a dreamy atmosphere in this recording. Chorus guitar samples could be picked up from this recording. Some parts of it could also be mixed into advertisement jingles.
    \item The music is a soft rock ballad with Hawaiian sounds. The melody is driven by a gliding steel pedal guitar. In the background a synth string section can be heard sustaining chords creating a relaxed ambience. The atmosphere of the song is easy going and relaxed. This is a song you would hear at a hotel lobby in Hawaii.
    \item Different melodies are being played by bell sounds;  e-piano; e-bass and in the background you can hear drone sounds and soft pads. A digital drum is playing a laid back groove. A sound-effect is coming in. The whole song sounds relaxing. This song may be playing for an advertisement.
    \item This song features the main melody played on a steel pan. This is accompanied by programmed percussion with the hi-hat being played on the 'and' count of every bar. A bass plays the root notes of the chords in the background. Another percussion instrument plays a beat that sounds like a trotting horse. This song has a happy mood. This song can be played in a movie where a family is going on a holiday at a Caribbean island.
    \item The rock music is purely instrumental and features an electric guitar with a distortion effect on. The guitarist is playing a lot of notes with virtuosity. This music excerpt would make for a really good electric guitar solo.
\end{enumerate}
\end{tiny}

\end{document}